\documentclass[prd,twocolumn,nofootinbib,amsfonts]{revtex4-2}
\usepackage[utf8]{inputenc}
\usepackage{amssymb}
\usepackage{amsmath}
 \usepackage{natbib}
\usepackage{graphicx}
\usepackage{dcolumn}
\usepackage{bm}
\usepackage{leftidx}
\usepackage{txfonts}
\usepackage{xcolor}
\usepackage{enumitem}
\usepackage{floatrow}
\usepackage{amsmath}
\usepackage{hyperref}
\RequirePackage{mathptmx}      
\RequirePackage{latexsym}
\newcommand{\mycomment}[1]{}

\begin{document}
\title{Null integrability and photon phenomenology in the accelerated Schwarzschild black hole}

\author{Shokoufe Faraji$^{1,2,3,4}$}
\email{s3faraji@uwaterloo.ca}
\author{Eva Hackmann$^{4}$}
\email{eva.hackmann@zarm.uni-bremen.de}

\affiliation{$^{1}$Department of Physics and Astronomy, University of Waterloo, Waterloo, N2L 3G1 Canada}
\affiliation{$^{2}$Waterloo Centre for Astrophysics, University of Waterloo, Waterloo, N2L 3G1 Canada}

\affiliation{$^{3}$Perimeter Institute for Theoretical Physics,  Waterloo, N2L 2Y5, Canada}
\affiliation{$^{4}$Center of Applied Space Technology and Microgravity (ZARM), University of Bremen, Am Fallturm 2, 28359 Germany}

\begin{abstract}
Uniformly accelerated black holes provide a controlled setting to isolate how translatory acceleration and conical defects modify photon dynamics and observable optics. We analyse null geodesics in the subextremal static C-metric, describing an accelerated Schwarzschild black hole pulled by a cosmic string/strut, and we keep the conicity parameter explicit throughout. By using dimensionless variables and a Mino-type parametrisation, the conformal Hamilton-Jacobi equation separates and both the radial and polar sectors reduce to Biermann-Weierstrass form, yielding compact closed solutions in terms of Weierstrass elliptic functions. This framework enables a systematic classification of photon trajectories, exhibits the loss of equatorial symmetry for nonzero acceleration via a fixed photon cone of constant-latitude null motion, and identifies a unique spherical photon surface shared by all latitudes. On the observational side, using the tetrad screen map, the circular shadow boundary follows transparently from the photon surface condition. We highlight the explicit cancellation of the conicity parameter and discuss that the angular radius depends on the acceleration and the observer's position; however, it is independent of the observer's inclination and of the conicity (because of the string tension). For nonstatic observers, a local Lorentz boost preserves circularity while changing the apparent angular size through aberration. We describe an algebraic inversion that infers the acceleration from a single shadow radius measurement once the mass-distance scale is fixed. Moreover, we derive closed expressions for the photon orbital frequency and Lyapunov exponent, providing Eikonal quasinormal mode estimates. These results supply a fully analytic toolkit for ray tracing and parameter inference in accelerated black hole spacetimes, and clarify which observables are (and are not) sensitive to conical defects.



\end{abstract}

\maketitle


\section{Introduction}
Black holes undergoing uniform translatory acceleration are modelled by the static C-metric. This spacetime is a well known solution to Einstein's equation that describes Schwarzschild geometry attached to a string/strut that supplies the acceleration and introduces a conical deficit along the symmetry axis \cite{1983GReGr..15..535B,2003CQGra..20.3269H,2006IJMPD..15..335G,2006CQGra..23.6745G}, and has been studied in many disciplines, e.g., \cite{PhysRevD.103.024007,2017JHEP...05..116A,2019PhLB..796..191G,2021CQGra..38s5024B,2021FrP.....9..187G,2021PhRvD.104d6007F,2023PhRvL.130i1603B,2025CQGra..42l5007L,2025PhRvD.111j4004H}. 
The spacetime originally belongs to the large class of solutions discovered by \cite{1918NCim...16..105L} and has a double horizon structure inside a static patch: the usual black hole horizon and an outer acceleration horizon. Beyond its exactness and conceptual clarity, this geometry provides a controlled setting in which to isolate how acceleration and conical defects alter geodesic motion and observables \cite{Morris2017,Chudnovsky1986} by primordial black holes \cite{Vilenkin2018}.

Cosmic strings appear in many scenarios describing the early universe, and might create pairs of black holes by breaking \cite{Hawking1995}, or be  dragged into galaxies by primordial black holes. They might appear as filaments attached to supermassive black holes e.g., \cite{Chudnovsky1986}, and recently viable detection channels of the resulting black hole acceleration have attracted considerable attention \cite{Ashoorioon2023}.

In the static C-metric the null Hamilton-Jacobi equation separates, and null orbits admit analytic treatment in terms of elliptic functions. Refs.~\cite{2001CQGra..18.1205P,Alawadi_2020,Frost2021,PhysRevD.103.024007} gave a detailed analysis of null and also timelike geodesics, including special orbits and integrability structures; later work extended the treatment to considering cosmological constant \cite{PhysRevD.103.024007} and exhibited explicit solutions in terms of Jacobi elliptic functions \cite{Frost2021}. Also, \cite{2015IJMPD..2442024G,2016PhLB..763..169G,2021PhRvD.103b5005Z} discussed photon surfaces and the shadow, and provided the characteristic radius relation used in later work. 

In this paper, null integrability refers to Liouville integrability of the
null geodesic flow: besides the null Hamiltonian constraint, the system admits
three additional independent first integrals in involution, so that the
massless Hamilton-Jacobi equation separates and the equations of motion reduce to
first order quadratures. In the static C-metric the two manifest integrals are the
energy and the axial angular momentum associated with stationarity and axisymmetry. The remaining separation constant is a Carter type integral generated by a
rank-2 conformal Killing tensor (equivalently, by an underlying conformal
Killing-Yano structure of the type~D Pleba\'nski-Demia\'nski class).
Because null geodesics are conformally invariant, this conformal hidden symmetry is
sufficient to guarantee complete integrability for photons, even though it does not
generically provide an additional constant for massive geodesics. For general background on hidden symmetries and their relation to separability and
integrability, see e.g.\ \cite{FrolovKrtousKubiznak:2017LRR,KrtousKubiznakPageVasudevan:2007GeodesicConstants}.
For the Pleba\'nski-Demia\'nski family (including accelerated black holes) and its
conformal Killing-Yano structure, see \cite{WalkerPenrose:1970QuadraticIntegralsTypeD,PlebanskiDemianski:1976PD,KubiznakKrtous:2007CKY_PD,2006IJMPD..15..335G}.
A recent viewpoint on null integrability in type~D backgrounds is given in
\cite{Andersson:2026MasslessSpinIntegrability}; in the spinless limit it reduces to the
standard integrability statement for null geodesics that we exploit here.

We recast the null problem so that both radial and polar equations reduce to the Biermann-Weierstrass form with explicit invariants following the method in \cite{Cieslik2022}. This yields particularly compact and manifestly real expressions for the solutions and for the invariants, without the need to calculate any turning points. This is an advantage compared to other works which typically use Jacobi elliptics or work with the standard Weierstrass form without normalization. Further, we recover the photon cone angle and the photon surface radius \cite{Alawadi_2020} as well as the associated constants of motion \cite{Frost2021} in consistent dimensionless notation and show transparently how the broken equatorial symmetry is encoded by the single radius together with the cone latitude.

For static (i.e. co-accelerated) observers, we show that the celestial coordinates of null geodesics generally depend on the conicity parameter. However, for any observer we prove that the shadow boundary is an exact circle, with a radius that is independent of conicity, generalising the result in \cite{2015IJMPD..2442024G,Frost2021}. We also show that the defining condition for the spherical photon surface\footnote{Here spherical just refers to a constant radius coordinate; as shown in \cite{2016PhLB..763..169G} by using an isometric embedding into euclidean space, the photon surface topology is actually not spherical.} can be reduced to a simple algebraic (quadratic) relation between the photon surface radius and the acceleration parameter; in particular, it is valid independently of the latitude and provides a convenient starting point for parameter inference from the shadow.

Many prior detailed studies of circular photon motion in the C-metric focused primarily on the equatorial case\footnote{Note that geodesics generally cannot be constrained to the equatorial plane of the accelerated black hole.} or on specific families. Here we instead emphasise a concise algebraic characterisation of the relevant circular orbit in the accelerated Schwarzschild case: constant-latitude motion is fixed by the photon cone, while the spherical photon orbit sits on a unique photon surface, and the associated conserved quantities follow in a uniform dimensionless normalisation.

 
 In addition, applying the Cardoso \textit{et al.}\ correspondence \cite{PhysRevD.79.064016}, we derive compact expressions for the orbital frequency and the Lyapunov exponent of the photon circle\footnote{Following \cite{2019PhRvD.100b4018G}, we use ``photon ring'' to refer to the bright image feature on the observer's screen, not to the underlying null geodesic. We refer to the orbit itself as a spherical photon orbit or photon circle in the case of constant latitude}. giving immediate quasinormal mode (QNM) eikonal estimates specialised to the accelerated Schwarzschild case.

Earlier analyses identified fixed-latitude photon motion (photon cones) \cite{Alawadi_2020}, established the existence and topology of photon surfaces in the C-metric \cite{2015IJMPD..2442024G,2016PhLB..763..169G,Frost2021,2021PhRvD.103b5005Z}, demonstrated shadow circularity for specific conventions or choices of the conicity parameter (e.g.\ \cite{PhysRevD.103.024007,2015IJMPD..2442024G,Frost2021}), often working with Jacobi elliptic functions or without a consistent dimensionless normalisation, and numerically calculated QNM \cite{2020PhRvD.102d4005D}. Our contribution is a streamlined, fully dimensionless and parameter-consistent treatment that (a) puts the null Hamilton-Jacobi separation into Biermann-Weierstrass form with explicit invariants for both sectors, (b) proves circularity of the shadow for any observer and shows why the angular deficit drops out of the local boundary while still affecting global lensing, (c) provides an algebraic inversion from a single shadow radius to the acceleration parameter, and (d) supplies compact, closed expressions for the photon circle frequency and instability that specialise the Cardoso \textit{et al.}\ correspondence to the accelerated Schwarzschild case \cite{PhysRevD.79.064016,2020PhRvD.102d4005D}. Building on earlier analyses of photon regions, shadows, and lensing in the C-metric, we keep the conicity parameter explicit and show how it enters the local screen
coordinates while cancelling from the boundary radius of the shadow.
This isolates what is genuinely local (the circular capture boundary) from what is
global (lensing and image multiplicities controlled by the angular deficit).

The paper is organized as follows: Section \ref{sec:metric} reviews the metric, horizons, and constants of motion. Section \ref{sec:null-dimless} develops the dimensionless null geodesic equations and their Biermann-Weierstrass solutions, while Section \ref{sec:photon_properties} discusses the photon cone and photon surface. Section \ref{sec:astro} turns to observables: the circular shadow radius for any observers, the (non)role of the conical deficit in the local shadow, a simple inversion for acceleration, as well as photon circle dynamics and eikonal ringdown. Section \ref{subsec:diagnostics} presents the surface gravities/temperatures of both horizons, and the redshift map between static worldlines. We conclude with a summary in Section \ref{sec:conclusions}. Throughout the paper we work in geometric units $G=c=1$ and Lorentzian signature $(-,+,+,+)$. Greek indices run over $(t,r,\theta,\phi)$.

\section{Static C-Metric and Constants of Motion}
\label{sec:metric}


The static C-metric describes a Schwarzschild black hole
uniformly accelerated along the $z$-axis, the acceleration being
provided by a (generally unbalanced) cosmic string or strut on the symmetry axis. In Boyer-Lindquist type of coordinates $(t,r,\theta,\phi)$, its Lorentzian section is 

\begin{equation}
ds^2= \frac{1}{\Omega^{2}}
\,\left[
 -\,\frac{\Delta}{r^{2}}\,dt^{2}
 +\,\frac{r^{2}}{\Delta}\,dr^{2}
 +\,\frac{r^{2}}{P}\,d\theta^{2}
 +\,P\,r^{2}\sin^{2}\theta\,d\phi^{2}
\right]
\label{eq:Cmetric}
\end{equation}
with
\begin{subequations}
\begin{align}\label{metriccomponents}
&\Omega(r,\theta) = 1+\alpha r\cos\theta,\\
&\Delta(r) = \bigl(r^{2}-2mr\bigr)\,\bigl(1-\alpha^{2}r^{2}\bigr),\\
&P(\theta) = 1+2\alpha m\cos\theta .
\end{align}
\end{subequations}
Here, $m$ is the black hole mass and $\alpha$ the (constant) proper acceleration. The functions $\Delta(r)$ and $P(\theta)$ are strictly positive in the "static patch"
\begin{equation}
0\;\le\;\alpha<\frac{1}{2m},\qquad
2m < r < \frac{1}{\alpha},\qquad
0<\theta<\pi,
\label{eq:static_region}
\end{equation}
so that $g_{tt}<0$, $g_{rr}>0$, $g_{\theta\theta}>0$, $g_{\phi\phi}>0$. Moreover,
$\Omega>0$ throughout the patch since
$\Omega_{\rm min}=1-\alpha r>0$ (attained at $\cos\theta=-1$ and
$r<1/\alpha$).


Equation \eqref{eq:Cmetric} is locally regular but may possess conical singularities on the symmetry axes ($\theta=0,\pi$). To encode a possible imbalance one allows the azimuthal coordinate to cover the range $-\pi C<\phi<\pi C$, where the conicity parameter $C>0$ is constant. After rescaling $\phi \mapsto C\,\phi$ the metric retains the form Equation \eqref{eq:Cmetric} but with $g_{\phi\phi}\rightarrow P\,C^{2}r^{2}\sin^{2}\theta$.

Near $\theta=0$ and $\theta=\pi$ the ratio of circumference to radius yields the (unsigned) conical deficits
\begin{equation*}
\delta_{\rm N}=2\pi\,\bigl[1-C(1+2\alpha m)\bigr],\quad
\delta_{\rm S}=2\pi\,\bigl[1-C(1-2\alpha m)\bigr].
\end{equation*}
Choosing $C_{\rm N}=1/(1+2\alpha m)$ regularizes the north axis and leaves a cosmic string of tension $\mu_{\rm S}=\delta_{\rm S}/(8\pi)$ on the south axis; conversely $C_{\rm S}=1/(1-2\alpha m)$ regularizes the south axis. We keep $C$ explicit to be able to trace its implications in detail; e.g., it enters the impact parameter of photons and the axial angular momentum of massive particles.


Regarding the horizons, the stationary Killing field $\xi_{(t)}=\partial_t$ becomes null at $g_{tt}=0$, i.e.\ at the zeros of $\Delta(r)$,
\begin{equation}
\Delta(r)=(r^{2}-2mr)\,(1-\alpha^{2}r^{2})=0\,.
\end{equation}
The roots are $r_{\rm H}=2m$, the Schwarzschild (black hole) horizon, and $r_{\rm A}=1/\alpha$, the acceleration horizon. The condition $\alpha<1/(2m)$ guarantees $r_{\rm H}<r_{\rm A}$, so the region $2m<r<1/\alpha$ is static.

\subsection{Killing fields and conserved quantities}
The C-metric, Equation \eqref{eq:Cmetric}, is stationary and axisymmetric, admitting the Killing vectors
\begin{equation}
\xi_{(t)}=\partial_{t},\qquad
\xi_{(\phi)}=\partial_{\phi}.
\end{equation}
For a geodesic $x^\mu(s)$ with $u^\mu=dx^\mu/ds$ the associated first
integrals are
\begin{subequations}
\begin{align}
E&:= -\,\xi_{(t)}\,\cdot\,u
     = -g_{tt}\,\dot t
     = \frac{\Delta}{\Omega^{2}r^{2}}\;\dot t,
\\
L&:=  \xi_{(\phi)}\,\cdot\,u
     =  g_{\phi\phi}\,\dot\phi
     = \frac{P\,C^{2}r^{2}\sin^{2}\theta}{\Omega^{2}}\;\dot\phi,
\label{eq:ELdefs_final}
\end{align}
\end{subequations}
with $E>0$ for future directed motion. The normalization
$g_{\mu\nu}u^{\mu}u^{\nu}=\delta$ distinguishes timelike ($\delta=-1$) and null ($\delta=0$) trajectories.

Additionally, in Hamiltonian form, with canonical momenta $p_\mu=g_{\mu\nu}\dot x^\nu$
(dots denote differentiation with respect to an affine parameter), geodesic
motion is generated by the conserved Hamiltonian
\begin{equation}
\mathcal H=\frac12\,g^{\mu\nu}p_\mu p_\nu=\frac12\,\delta,
\end{equation}
where again $\delta=0$ for null and $\delta=-1$ for timelike geodesics.
For null geodesics, in addition to the Killing integrals
$E=-p_t$ and $L=p_\phi$, there exists a further independent constant of motion
of Carter type associated with a rank-2 conformal Killing tensor:
\begin{equation}
K=\mathcal K^{\mu\nu}p_\mu p_\nu, 
\label{eq:CKT_and_CarterK}
\end{equation}
where 
\begin{equation}
\mathcal K^{\mu\nu}=\frac{1}{P(\theta)}
\left(\delta^\mu_{\ \theta}\delta^\nu_{\ \theta}
+\frac{1}{C^{2}\sin^{2}\theta}\,\delta^\mu_{\ \phi}\delta^\nu_{\ \phi}\right),
\label{eq:CKT_and_CarterK2}
\end{equation}
and $\delta^\mu_{\ a}$ refers to the Kronecker symbol. Equivalently,
\begin{equation}
K=\frac{1}{P(\theta)}\left(p_\theta^{2}+\frac{p_\phi^{2}}{C^{2}\sin^{2}\theta}\right)\ge0.
\end{equation}
This constant coincides with the separation constant introduced in the
Hamilton-Jacobi analysis below. In the spherical Schwarzschild limit
($\alpha\to0$ and $C\to1$) it reduces to the usual total angular momentum squared
$L_{\rm tot}^{2}=p_\theta^{2}+p_\phi^{2}/\sin^{2}\theta$. Such hidden symmetry structures (conformal Killing -Yano/Killing tensors and
Carter type integrals) are standard in type~D families including the
(Pleba\'nski -Demia\'nski) C-metric; see, e.g., \cite{KubiznakKrtous:2007CKY_PD,2014PhRvD..89j4016L,WalkerPenrose:1970QuadraticIntegralsTypeD,Carter1968}.

\section{Null geodesics in dimensionless variables}
\label{sec:null-dimless}

Throughout this section we specialise to null geodesics
($g_{\mu\nu}u^\mu u^\nu=0$) and convert all quantities to
dimensionless form by scaling with the black hole mass $m$. For null geodesics the affine parameter can be rescaled by a constant, so $(E,L,K)$ have no absolute normalisation; throughout we fix this freedom once and work with the affine--invariant ratios introduced below:

\begin{equation}
\xi := \frac{r}{m},\,
\tau := \frac{t}{m},\,
a := \alpha m,\,
\varepsilon := E,\,
\ell := \frac{L}{m},\,
\ell_{C} := \frac{\ell}{C},
\label{eq:dimless_defs}
\end{equation}
where $a = \alpha m \in (0,\tfrac12)$. The metric functions are
\begin{subequations}
\begin{align}
&\Omega(\xi,\theta)=1+a\,\xi\cos\theta,\\
&\Delta(\xi)=\xi^{2}\Bigl(1-\tfrac{2}{\xi}\Bigr)\bigl(1a^{2}\xi^{2}\bigr),\label{eq:Delta_dimless}\\
&P(\theta)=1+2a\cos\theta,
\label{eq:dimless_functions}
\end{align}
\end{subequations}
and the static patch is $2<\xi<1/a$ for $a>0$ (with $\xi>2$ when $a=0$), $0<\theta<\pi$.


\subsection{Mino-type parameter, Hamilton-Jacobi separation, and $K$-scaling}
Because null motion is conformally invariant, we work with the conformal metric $\hat g_{\mu\nu}=\Omega^{2}g_{\mu\nu}$. Let $\tilde{\lambda}$ be an affine parameter for $\hat{g}_{\mu\nu}$.

With the additive ansatz
$S=-\varepsilon\,\tau+\ell_{C}\,\phi+S_{r}(\xi)+S_{\theta}(\theta)$,
the conformal Hamilton-Jacobi equation
$\hat g^{\mu\nu}\partial_\mu S\,\partial_\nu S=0$ separates,
\begin{subequations}\label{eq:HJsep_dimless}
\begin{align}
&\bigl(S_r'\bigr)^{2}
-\xi^{4}\,\varepsilon^{2}=- \Delta(\xi) K,
\label{eq:HJsep_r_dimless}\\
&\bigl(S_\theta'\bigr)^{2}
+\frac{\ell_{C}^{2}}{\sin^{2}\theta}=P(\theta) K,
\label{eq:HJsep_theta_dimless}
\end{align}
\end{subequations}
where primes denote ordinary derivatives and $K\ge0$ is a separation
constant (the analogue of Carter's constant in Kerr; nonnegativity
follows from Equation \eqref{eq:HJsep_theta_dimless} since both terms are nonnegative in the static patch). Introducing the dimensionless impact parameter

\begin{equation}
\beta:=\frac{\ell}{\varepsilon}=\frac{1}{m}\,\frac{L}{E},
\qquad
Q:=\frac{K}{\varepsilon^{2}},
\end{equation}
the first order null equations become


\begin{subequations}\label{eq:null_first_tildelam_dimless}
\begin{align}
&\left(\frac{d\xi}{d\tilde\lambda}\right)^{2}
=\varepsilon^{2}\left[1-\frac{\Delta(\xi)}{\xi^{4}}Q\right],\label{eq:dximinotilde}\\
&\left(\frac{d\theta}{d\tilde\lambda}\right)^{2}
=\frac{1}{\xi^{4}}\left[P(\theta)K-\frac{\ell_{C}^{2}}{\sin^{2}\theta}\right],
\label{eq:dthetaminotilde}\\
&\quad \frac{d\phi}{d\tilde\lambda}
=\frac{\ell_{C}}{P(\theta)C\xi^{2}\sin^{2}\theta}\\
&\quad \frac{d\tau}{d\tilde\lambda}=\frac{\xi^{2}\varepsilon}{\Delta(\xi)}.
\label{eq:phitautildelam}
\end{align}
\end{subequations}
Radial turning points satisfy $\xi^{4}=\Delta(\xi)\,Q$; angular turning points satisfy

\begin{equation}
Q=\frac{\beta^{2}}{C^{2}\,P(\theta)\,\sin^{2}\theta}.
\end{equation}

To completely decouple the above equations, we define a
Mino-type parameter $\lambda$ by\footnote{We choose the affine parameter $\tilde\lambda$ of the conformal metric $\hat g_{\mu\nu}$
to be dimensionless (any affine parameter may be rescaled by a constant).}
\begin{equation}
d\lambda=\frac{1}{\xi^{2}}\,d\tilde{\lambda} .
\label{eq:def_mino}
\end{equation}
 For later convenience, additionally we absorb $K>0$ by a constant reparametrisation and scaled constants:

\begin{equation}
\lambda_{K}:=\sqrt{K}\,\lambda,\qquad
\hat e:=\frac{\varepsilon}{\sqrt{K}},\qquad
\hat\ell:=\frac{\ell_{C}}{\sqrt{K}} .
\label{eq:K_scaling}
\end{equation}
The degenerate case $K=0$ forces (from Equation \eqref{eq:HJsep_theta_dimless})
$\ell_{C}=0$ and $S_\theta' = 0$, i.e. $\theta=\text{const}$ and
${d\phi}/{d\tilde\lambda}=0$: these are radial null rays at arbitrary fixed latitude. Such rays should be treated in the unscaled Mino system
by setting $K=0=\ell_{C}$ before the $K$-rescaling Equation \eqref{eq:K_scaling}. In the Schwarzschild limit $a=0$ the Carter-like constant $K$ reduced to the squared total angular momentum, and consequently $\hat \ell$ to the cosine of the inclination.

The first order equations become

\begin{subequations}\label{eq:firstorder_dimless}
\begin{align}
&\left(\frac{d\xi}{d\lambda_{K}}\right)^{2}
= \xi^{4}\,\hat e^{2}-\Delta(\xi),
\label{eq:first_r}\\[3pt]
&\left(\frac{d\theta}{d\lambda_{K}}\right)^{2}
= \,P(\theta)-\frac{\hat\ell^{2}}{\sin^{2}\theta},
\label{eq:first_theta}\\
&\quad \frac{d\phi}{d\lambda_{K}}
=\frac{\hat\ell}{CP(\theta)\,\sin^{2}\theta},\\
&\quad\frac{d\tau}{d\lambda_{K}}
=\frac{\xi^{4}\,\hat e}{\Delta(\xi)}.
\label{eq:first_phi_tau}
\end{align}
\end{subequations}
Radial turning points are the positive roots of
$\xi^{4}\hat e^{2}-\Delta(\xi)=0$, and polar motion is allowed where
\begin{equation}
P(\theta)-\frac{\hat\ell^{2}}{\sin^{2}\theta}\;\ge 0\;
\quad\Longleftrightarrow\quad
\sin^{2}\theta\;\ge\;\frac{\hat\ell^{2}}{P(\theta)}.
\end{equation}
Unless $a=0$, the reflection symmetry
$\theta\mapsto\pi-\theta$ is broken.

\subsection{Radial quartic and Weierstrass reduction}
From Equation ~\eqref{eq:first_r} we define
\begin{equation}
R(\xi):=\xi^{4}\hat e^{2}-\Delta(\xi)
\end{equation}
with $\Delta(\xi)$ given in Equation~\eqref{eq:Delta_dimless}. Expanding $R(\xi)$ yields the explicit quartic
\begin{equation}
R(\xi)
=\bigl(\hat e^{2}+a^{2}\bigr)\,\xi^{4}-2a^{2}\,\xi^{3}
-\xi^{2}+2\xi .
\label{eq:Rquartic}
\end{equation}
Zeros of $R(\xi)$ determine radial turning points. Following \cite{Cieslik2022,Cieslik2023}, see also the appendix \ref{app:null:weierstrass}, we solve the radial equation using the Biermann-Weierstrass approach. For any initial value $\xi_0$ the solution then reads
\begin{equation}
\xi(\lambda_K) = \xi_0 + \Pi,
\end{equation}
where
\begin{equation}
\Pi = \frac{R_0'(\wp(\lambda_K)-\frac{1}{24}R''_0)-2\epsilon_r\sqrt{R_0}\wp'(\lambda_K)+\frac{1}{12}R_0R_0'''}{4(\wp(\lambda_K)-\frac{1}{24}R_0'')^2 - \frac{1}{24}R_0R_0''''}\,.\label{eq:solr}
\end{equation}
Here the index $0$ indicates evaluation at the initial value $\xi_0$, e.g. $R_0' = R'(\xi_0)$, and the primes indicate a derivative of $R$ with respect to $\xi$. The symbol $\epsilon_r = \pm 1$ encodes the initial propagation direction: $\epsilon_r=+1$ ($\epsilon_r=-1$) for initially increasing (decreasing) radius. The Weierstrass function $\wp(\lambda_K) = \wp(\lambda_K;g_2,g_3)$ depends on the invariants $g_2$, $g_3$ that are independent of the chosen initial value. The general expression for a generic quartic can be found in the appendix, ~Equation \eqref{app:g2g3}. For the radial case at hand they are

\begin{subequations}
    \begin{align}
&g_{2}^{(r)}=\frac{1}{12}+a^{2},\\
&g_{3}^{(r)}=\frac{1}{216}-\frac{a^{2}}{6}-\frac{\hat e^{2}}{4}.
\label{eq:Weier_r_Kscaled}
\end{align}
\end{subequations}
Because of the $K$-scaling, these invariants are $K$ independent.
In the Schwarzschild case $a=0$ this reduces to the known expressions \cite{Cieslik2022}. 

The manifestly real general solution above has the advantage that we do not need to solve the quartic to find any turning points; knowledge of the K-scaled energy is sufficient. However, for the particular case that $\xi_0$ is a turning point, $R(\xi_0)=R_0=0$, the general solution \eqref{eq:solr} simplifies significantly to
\begin{equation}
 \Pi = \frac{R_0'}{4\wp(\lambda_K)-R_0''/6}\,. \label{eq:xi_r0_solution}   
\end{equation}
Since $R(0)=0$, choosing the formal initial value $\xi_0=0$ (outside the static patch) rewrites the solution in the compact form

\begin{equation}
    \xi(\lambda_K)= \frac{1}{2\wp(\lambda_K) + \frac{1}{6}}\,.
\end{equation}
In addition, spherical photon orbits occur when $R(\xi)$ has a positive double root, i.e.\ $R(\xi_{\rm ph})=R'(\xi_{\rm ph})=0$, which yields the known
radius $\xi_{\rm ph}(a)=6/(1+\sqrt{1+12a^{2}})$ used later. 



\subsection{Polar equation and Weierstrass reduction}
Let $\nu:=\cos\theta$, so that $P(\theta)=1+2a\,\nu$ and
$\sin^{2}\theta=1-\nu^{2}$. From Equation \eqref{eq:first_theta} and using $d\nu/d\lambda_{K}=-\sin\theta\,d\theta/d\lambda_{K}$ we obtain
\begin{equation}
\left(\frac{d\nu}{d\lambda_{K}}\right)^{2}
= (1+2a\,\nu)\,(1-\nu^{2})-\hat\ell^{2} =: \Theta_\nu(\nu).
\label{eq:polar_nu_lamK}
\end{equation}
Turning latitudes satisfy
\begin{equation*}
(1+2a\,\nu_{\,*})\,(1-\nu_{\,*}^{2})=\hat\ell^{2}\,,
\end{equation*}
i.e. $P(\theta_{\,*})\,\sin^{2}\,\theta_{\,*}=\hat\ell^{2}$. Since $0<a<\tfrac12$ one has $P(\theta)>0$ on $(0,\pi)$, so polar motion is allowed precisely where the right hand side of Equation \eqref{eq:polar_nu_lamK} is non‑negative. 

The Biermann-Weierstrass solution of Equation \eqref{eq:polar_nu_lamK} is analogous to the radial case. We find

\begin{equation}
\nu(\lambda_K) = \nu_0 + \chi,\label{eq:solnu}
\end{equation}
where
\begin{equation*}
\chi = \frac{\Theta_0'(\wp(\lambda_K)-\frac{1}{24}\Theta''_0)-2\epsilon_\nu\sqrt{\Theta_0}\wp'(\lambda_K)+\frac{1}{12}\Theta_0\Theta_0'''}{4(\wp(\lambda_K)-\frac{1}{24}\Theta_0'')^2 - \frac{1}{24}\Theta_0\Theta_0''''}.
\end{equation*}
Here $\nu_0$ is the initial value of $\nu=\cos\theta$, primes are derivatives with respect to $\nu$, and an index $0$ indicates evaluation at the initial value, e.g. $\Theta_0'=\Theta'_\nu(\nu_0)$. As in the radial case, $\epsilon_\nu=+1$ ($\epsilon_\nu = -1$) for initially increasing (decreasing) $\cos\theta$. The invariants of the Weierstrass $\wp$ function are

\begin{subequations}
\begin{align}
&g_{2}^{(r)}=\frac{1}{12}+a^{2},\label{eq:g2_r}\\
&g_{3}^{(\theta)}=\frac{1}{216}
+a^{2}\,\left(\frac{\hat\ell^{2}}{4}-\frac{1}{6}\right).
\label{eq:g2g3_theta_final}
\end{align}    
\end{subequations}
Notice $g_{2}^{(\theta)}$ coincides with the radial $\,g_{2}^{(r)}\,$,
cf. Equation \eqref{eq:g2_r}. As before, the general equation simplifies significantly, if we choose a turning point as the initial value. In this case,
\begin{equation}
    \chi =  \frac{\Theta_0'}{4\wp(\lambda_K) - \Theta_0''/6}\,. \label{eq:nu_solution}
\end{equation}

\subsection{Azimuth and coordinate time as Weierstrass quadratures}

The remaining first integrals in Equation ~\eqref{eq:first_phi_tau} involve
rational functions of $\xi$ or $\theta$. After substituting the
Weierstrass solutions for the polar and radial sectors they reduce to integrals of the type $\int dz\,(\wp(z)-c)^{-1}$. We use the standard antiderivative: 

\begin{align}\label{eq:Weier_primitive}
\mathcal{F}^{(\,\cdot\,)}_{c}(z)
&:=\int^{z}\frac{dw}{\wp(w;g_{2}^{(\,\cdot\,)},g_{3}^{(\,\cdot\,)})-c}\\
&=\frac{1}{\wp'\,\bigl(z_{c}^{(\,\cdot\,)}\bigr)}
\left[
\log\frac{\sigma\bigl(z-z_{c}^{(\,\cdot\,)}\bigr)}{\sigma\bigl(z+z_{c}^{(\,\cdot\,)}\bigr)}
+2z\,\zeta\,\bigl(z_{c}^{(\,\cdot\,)}\bigr)
\right]\nonumber,
\end{align}
where we dropped the invariants in the second line and $z_{c}^{(\,\cdot\,)}$ is any preimage satisfying
$\wp\,\bigl(z_{c}^{(\,\cdot\,)};g_{2}^{(\,\cdot\,)},g_{3}^{(\,\cdot\,)}\bigr)=c$
(the result is independent of that choice up to periods).

\begin{figure}
\centering
\includegraphics[width=0.47\textwidth]{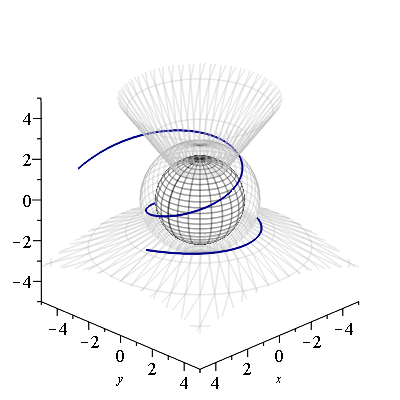}\quad
\includegraphics[width=0.47\textwidth]{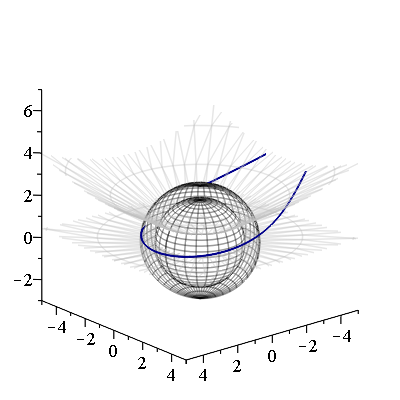}
\caption{Null trajectories in the static C-metric for $a=0.375$ and $C=1$. Left: Region I (equator‑crossing) example with $\hat{e}=\varepsilon/\sqrt{K}=\,0.07$ and $\hat{\ell}=\ell/\sqrt{K}=\,0.75$ ($\hat{\ell}^2=0.5625$). 
Right: Region II (hemisphere‑confined) example with $\hat{e}=\varepsilon/\sqrt{K}=0.05$ and $\hat{\ell}=\ell/\sqrt{K}=\,1.02$ ($\hat{\ell}^2=\,1.0404$), and $\hat{e}^2_{\rm crit}\,=\Delta(\xi_{\rm ph})/\xi_{\rm ph}^4 \, \approx 6.37 \times 10^{-3}$. In both panels, $\hat{e}^2_{\rm crit} \leq \hat{e}^2$, hence the radial motion has a single turning point located near $\xi_{\rm ph} \approx 2.273$. The black sphere marks the black hole horizon at $\xi=2$. a faint plane or ring at $\xi_{A}=1/a \approx 2.667$ for acceleration horizon. Translucent cones show the polar turning latitudes obtained from $P(\theta)\sin^{2}\theta=\hat{\ell}^2$. The left trajectory crosses the equator; the right trajectory remains entirely in the northern hemisphere. Coordinates: $(x,y,z)=\big( \xi \sin\theta \cos\phi,\xi \sin\theta \sin\varphi,\xi \cos\theta \big)$.}
\label{Fig:flybyorbits}
\end{figure}

\begin{figure}
\centering
\includegraphics[width=0.7\textwidth]{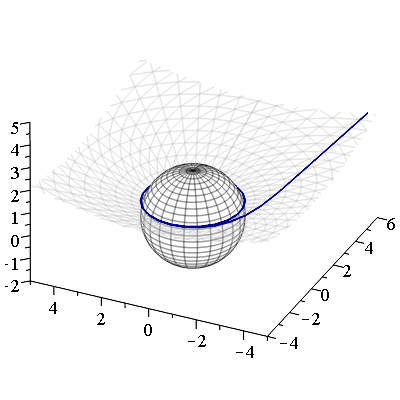}
\caption{Orbit which comes from infinity and spirals on an off-equatorial circular orbit. The radius of the circular orbit was chosen as $\xi=2.1$ (i.e.\ $r=2.1\,m$). It lies on a cone with $\cos\theta\approx 0.3$. }
\label{Fig:inspiral}
\end{figure}

\medskip

\paragraph*{(i) Azimuth $\phi(\lambda_{K})$.}
A partial fraction decomposition of the azimuthal equation reads
\begin{align}
    \frac{d\phi}{d\lambda_K} = \frac{\hat \ell}{C} \sum_{i=1}^3 \frac{\beta_i}{\nu-c_i} \,,
\end{align}
where
\begin{equation}
    c_1 = 1, \quad c_2 = -1, \quad c_3 = -\frac{1}{2a},
\end{equation}
and
\begin{equation}
   \beta_1 = \frac{-1}{2+4a}, \quad \beta_2 = \frac{1}{2-4a}, \quad \beta_3 = \frac{2a}{(4a^2-1)} \,.
\end{equation}
The identities \eqref{eq:xi_r0_solution} and \eqref{eq:nu_solution} allow one to rewrite each term $(\nu-c_i)^{-1}$ as a Weierstrass third-kind integrand by direct algebra (see Appendix~\ref{app:null:weierstrass}) as

\begin{equation}
\frac{1}{\nu-c_i} = \frac{1}{\nu_0-c_i} + \frac{\frac{1}{4}P_i'}{\wp(\lambda_K)-\frac{1}{24}P_i''} ,   
\end{equation}
assuming that $\nu_0$ is a turning point, and $P_i' = -\Theta_0'/(\nu_0-c_i)^2$, $P_i'' = \Theta_0'' - 6 \Theta_0'/(\nu_0-c_i)$. The solution for $\phi$ can then be written as
\begin{equation}
    \phi(\lambda_K) = \phi_0 + \frac{\hat \ell \lambda_K}{CP(\theta_0)\sin^2\theta_0} - \frac{\hat \ell}{C} \sum_{i=1}^3 \frac{ \Theta_0' \beta_i \mathcal{F}^{(\theta)}_{d_i}(\lambda_K)}{4(\nu_0-c_i)^2}, \label{eq:phi_solution} 
\end{equation}
where 

\begin{equation}
    d_i = \frac{P_i''}{24} = \frac{\Theta_0''}{24} - \frac{\Theta_0'}{4(\nu_0-c_i)},
\end{equation}
and each primitive uses the polar invariants and preimages, i.e. $\wp\,\bigl(z_{d_{i}};\,g_{2}^{(\theta)},g_{3}^{(\theta)}\bigr)=d_{i}$.

\medskip

\paragraph*{(ii) Coordinate time $\tau(\lambda_{K})$.}

We find for the radial sector
\begin{equation}
    \frac{d\tau}{d\lambda_K} = \frac{\hat{e}}{a^2} \left[ -1 + \sum_{i=1}^3 \frac{\gamma_i}{\xi-h_i} \right], 
\end{equation}

with coefficients
\begin{equation}
    \gamma_1 = \frac{1}{2a(2a-1)}, \quad \gamma_2 = \frac{1}{2a(2a+1)}, \quad \gamma_3 = \frac{8a^2}{1-4a^2},
\end{equation}
and $h_i$ given by the horizons
\begin{equation}
    h_1= \xi_A = \frac{1}{a}, \quad h_2 = -\xi_A, \quad h_3 = \xi_H = 2 \,.
\end{equation}
Assuming that $\xi_0$ is a turning point, we then obtain the solution for the dimensionless coordinate time $\tau=t/m$ as
\begin{equation}
    \tau(\lambda_K) = \tau_0 + \frac{\xi_0^{4}\hat{e}}{\Delta(\xi_0)} \lambda_K - \frac{\hat e}{a^2} \sum_{i=1}^3 \frac{R_0' \gamma_i \mathcal{F}^{(r)}_{H_i}(\lambda_K)}{4(\xi_0-h_i)^2} , \label{eq:tau_solution}
\end{equation}
where 
\begin{equation}
    H_i = \frac{R_0''}{24} - \frac{R_0'}{4(\xi_0-h_i)},
\end{equation}
and each primitive uses the radial invariants and preimages, i.e. 
$\wp\,\bigl(z_{H_{i}};\,g_{2}^{(r)},g_{3}^{(r)}\bigr)=H_{i}$.

Equations \eqref{eq:xi_r0_solution}, \eqref{eq:nu_solution}, \eqref{eq:phi_solution}, and \eqref{eq:tau_solution} provide a complete analytic solution for null geodesics in the static C-metric throughout the subextremal domain $a\in(0,\tfrac12)$.
In Figure \ref{Fig:flybyorbits}, we plotted two examples for a value of $a$ close to the upper bound $1/2$, to clearly show the impact of the acceleration parameter. The two horizons are very close together, and it is clearly visible how the flyby orbits deviate from plane motion. Also, Figure \ref{Fig:inspiral} shows an example of an orbit that comes from infinity and spirals on an off-equatorial circular orbit.


\medskip

In summary, qualitative structure and turning points are as follows:
\begin{enumerate}[label=\roman*)]

\item Radial motion. With $R(\xi):=\xi^{4}\hat e^{2}-\Delta(\xi)$ [Equation ~\eqref{eq:first_r}],
$R$ is a quartic with vanishing constant term ($R(0)=0$). 
Real, positive roots classify fly‑by, plunge, and spherical photon orbits. Spherical (constant–$\xi$) orbits occur at double roots: $R(\xi_{*})=0=R'(\xi_{*})$.

\item Polar motion. From Equation ~\eqref{eq:first_theta} and since $P(\theta)=1+2a\cos\theta>0$ for $0<a<\tfrac12$, motion is allowed precisely where $P(\theta)\sin^{2}\theta\ge \hat\ell^{2}$. Turning latitudes satisfy $P(\theta_{*})\,\sin^{2}\theta_{*}=\hat\ell^{2}$.
Because $P(\theta)$ is not symmetric under $\theta\mapsto\pi-\theta$ for $a\neq0$, the allowed latitude band is generally not symmetric about the equator.

\item Genus. The radial and polar motions obey independent Weierstrass equations, generally with different invariants. Therefore, each sector is elliptic (genus\,1), and the full null motion is quasi periodic on the 2-torus $\mathbb T^{2}$. 
The coordinates $\phi(\lambda_K)$ and $\tau(\lambda_K)$ are sums of elliptic (third kind) integrals of the two arguments. A single hyperelliptic (genus\,2) curve does not arise unless one eliminates $\lambda_K$ to express $\theta(\phi)$ or $r(\tau)$. If $\theta$ is constant (cone orbits at $\theta=\theta_{\gamma}(a)$) the polar phase is fixed and the quadratures reduce to a single Weierstrass argument; $\phi(\lambda_{K})$ is then linear in $\lambda_{K}$. Analogously, $\tau(\lambda_K)$ is linear in $\lambda_K$ for orbits with constant radius.

\end{enumerate}
\begin{figure}
\centering
\includegraphics[width=0.9\textwidth]{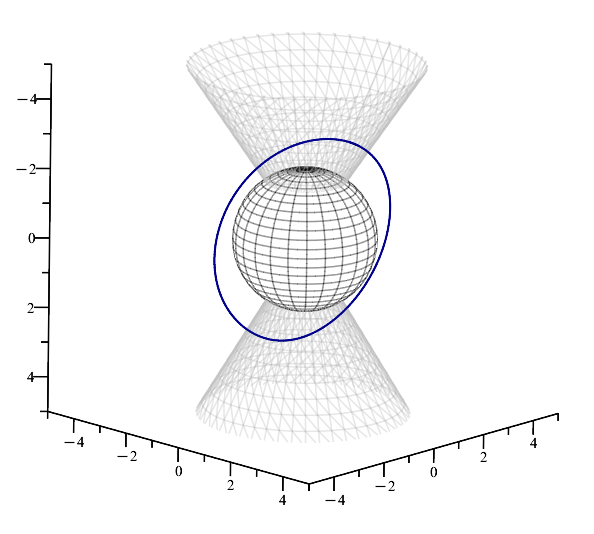}\quad
\includegraphics[width=0.8\textwidth]{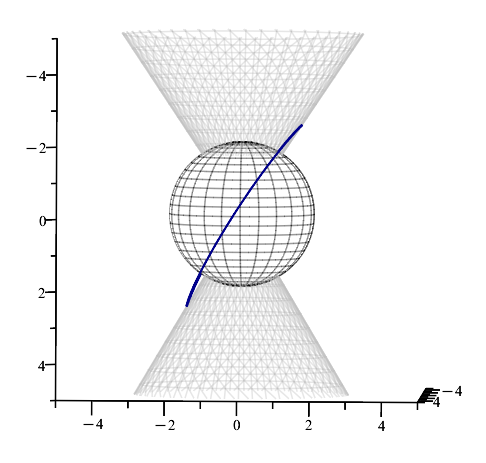}
\caption{Closed but non-circular null trajectory in the static C-metric (same orbit, two viewpoints). The blue curve shows a periodic photon orbit in the static patch, with $a=0.05$, $C=1$, $\hat e$ such that $\xi \equiv \xi_{\rm ph}(a)$ (see Equation \eqref{eq:ehat_ph_final}) and $\hat{\ell}$ such that the orbit closes, $\hat \ell \approx 0.531817$. The grey sphere marks the black hole horizon at $\xi=2$.
The translucent cones indicate the polar turning latitudes determined by the
$\theta$-sector turning condition (roots of the polar potential).
Unlike the circular photon orbit on the photon cone, this closed orbit is not confined to a single latitude and is not captured by planar intuition, reflecting the breakdown of equatorial symmetry for $a\neq0$.
Coordinates: $(x,y,z)=(\xi\sin\theta\cos\phi,\ \xi\sin\theta\sin\phi,\ \xi\cos\theta)$.}
\label{Fig:closed_non_circular}
\end{figure}

\begin{figure}
\centering
\includegraphics[width=0.8\textwidth]{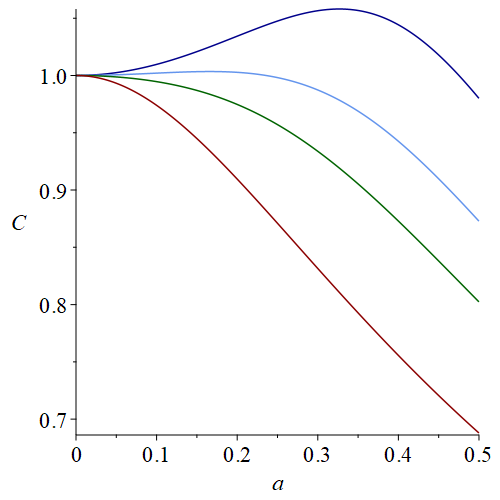}\quad
\caption{Closure (resonance) curves in the $(a,C)$ plane.
For each fixed angular momentum label (one curve per choice; red curve has $\hat \ell=0.9$, green $\hat \ell=0.6$, light blue $\hat \ell=0.5$, dark blue $\hat \ell = 0.4$), we compute the polar period $\Lambda_\theta$ of $\nu(\lambda_K)=\cos\theta$ and evaluate the azimuthal advance $\Delta\phi=\phi(\Lambda_\theta)-\phi(0)$ using the Weierstrass quadrature for $\phi(\lambda_K)$. The plotted curves show the values of the conicity parameter $C$ for which $\Delta\phi=2\pi$, i.e.\ the orbit closes after one polar oscillation (the simplest $1{:}1$ commensurability). Away from these curves the motion is quasi-periodic on $\mathbb{T}^2$.}
\label{fig:C_of_a_closure}

\end{figure}

\paragraph*{Closed orbits as commensurability curves in $(a,C)$.}
On $\xi=\mathrm{const}$ (i.e. on the photon surface $\xi=\xi_{\rm ph}$), the polar motion $\nu(\lambda_K)=\cos\theta(\lambda_K)$ is periodic with some Mino-time period $\Lambda_\theta$. The associated azimuthal advance over one polar period is
$\Delta\phi:=\phi(\Lambda_\theta)-\phi(0)$.
Generic trajectories are quasi-periodic on the 2--torus of $(\nu,\phi)$ phases; a closed curve occurs only when $\Delta\phi/(2\pi)\in\mathbb{Q}$.
In the simplest $1{:}1$ case we impose $\Delta\phi=2\pi$ and, for fixed angular momentum label, solve this condition for the conicity parameter $C$ as a function of the acceleration $a$.
We present a closed photon trajectory arising from the simplest $1{:}1$ ratio. 
Figure~\ref{Fig:closed_non_circular} shows a representative closed spherical orbit that is nevertheless not circular/planar:
the latitude oscillates between turning values with a varying rate different from the longitudinal $\phi$ advance. One can also explicitly show that the closed orbit is not planar by checking that the torsion of the spatial trajectory is nonvanishing, illustrating directly the loss of equatorial/planar orbit intuition when $a\neq0$.
Additionally, Figure~\ref{fig:C_of_a_closure} shows the resulting closure curves $C(a)$ (zoomed near $C\simeq1$), illustrating how the string/strut tension tunes the azimuthal precession relative to the polar oscillation.

\medskip




\section{Photon surfaces and orbital cones} \label{sec:photon_properties}

\begin{figure}
\centering
\includegraphics[width=0.46\textwidth]{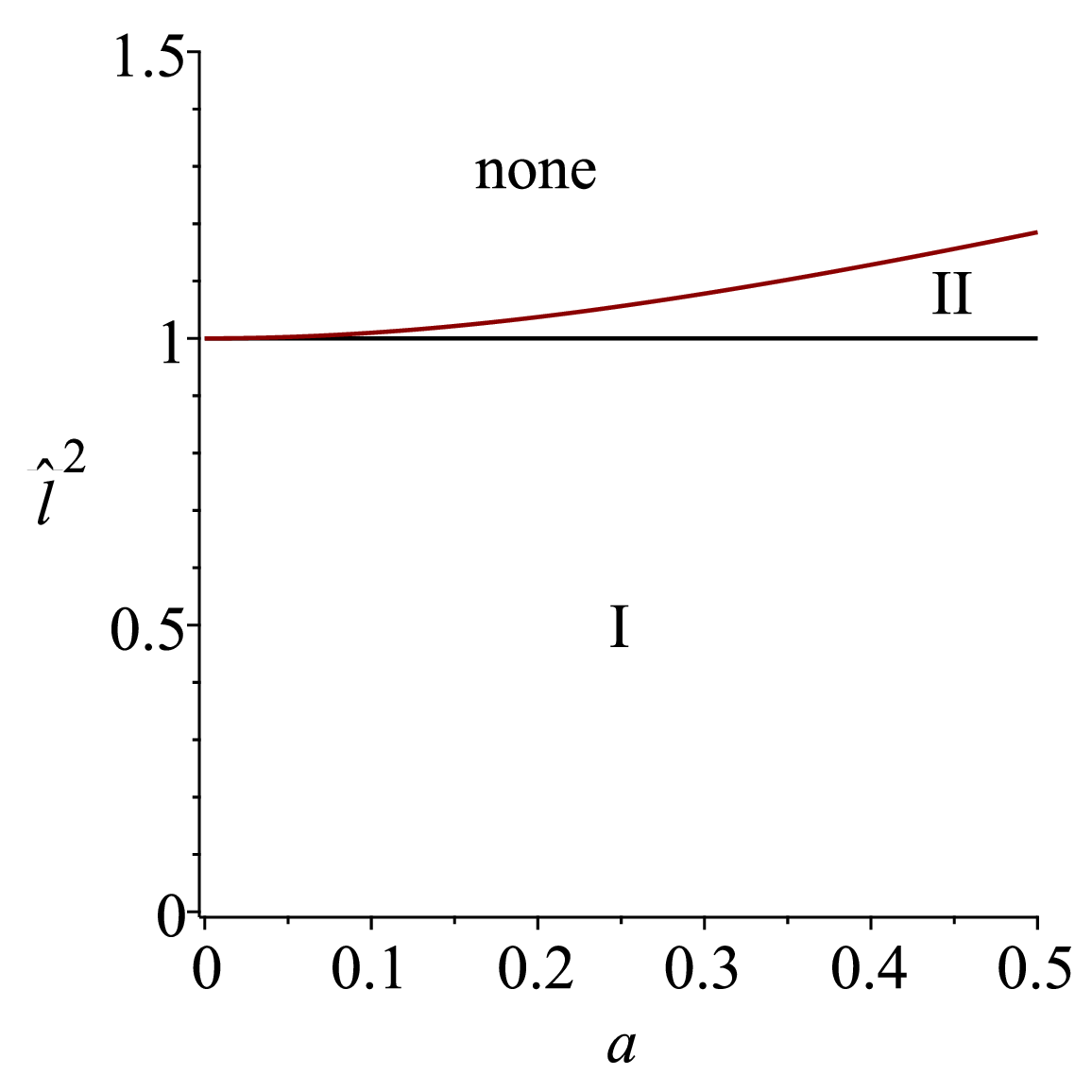}\qquad
\includegraphics[width=0.46\textwidth]{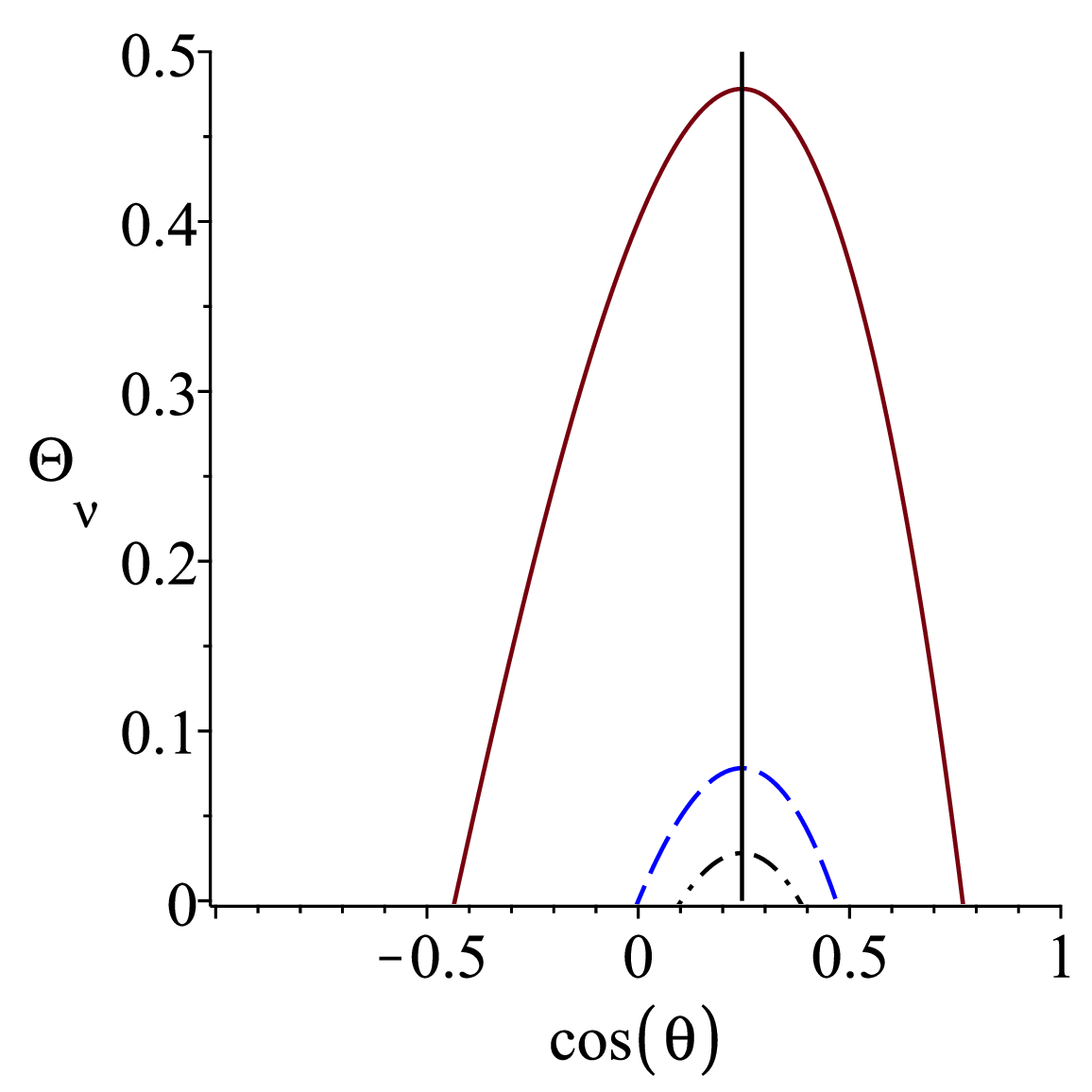}
\caption{Polar motion of null geodesics.
Left: classification in the plane $(a,\hat{\ell}^2)$. The black line is $\hat{\ell}=1$ (turning point at the equator). 
The red line is the fixed cone bound 
$\hat{\ell}^2=P(\theta_\gamma)\sin^2\theta_\gamma$ with $\cos\theta_\gamma=2a/[1+\sqrt{1+12a^2}]$ (see also Equation \eqref{eq:cone-angle}),  
above it no null geodesic motion is allowed. Region I ($\hat{\ell}<1$) oscillates about the equator; region II ($1<\hat{\ell}^2<P(\theta_\gamma)\sin^2\theta_\gamma$) is confined to one hemisphere. Right: $\Theta_\nu(\nu)$ vs $\nu=\cos\theta$ for $a=0.3$ and 
$\hat{\ell}\in\{0.6,1.0,1.05\}$. The vertical black line indicates the value of the photon cone $\cos\theta_\gamma$, that is an extremum of $\Theta_\nu$ regardless of the value of $\hat{\ell}$. }
\label{Fig:theta_Lvsalpha}
\end{figure}

\subsection{Photon cones and breakdown of equatorial symmetry}
\label{sec:photon-cone}
We continue with the $K$-scaled first integrals introduced above. In
particular, from Equation ~\eqref{eq:first_theta} we write
\begin{equation}
\Bigl(\frac{d\theta}{d\lambda_{K}}\Bigr)^{2}
=\Theta(\theta)
:=P(\theta)\,-\frac{\hat\ell^{2}}{\sin^{2}\theta}.
\label{eq:Theta_dimless_again}
\end{equation}
with $P(\theta)=1+2a\cos\theta$, $0<a<\tfrac12$.
A null geodesic remains at a fixed latitude $\theta=\theta_{\gamma}$ iff
(i) $\Theta(\theta_{\gamma})=0$ and (ii) the $\theta$ sector of Euler-Lagrange equation is satisfied at $d\theta/d\lambda_{K}=0$.

Condition (i) fixes the (scaled)
axial momentum to
\begin{equation}
\hat\ell^{2}=P(\theta_\gamma)\sin^{2}\theta_{\gamma}.
\label{eq:ellhat_cone}
\end{equation}
Therefore, the sign of $\hat\ell$ sets the azimuthal orientation: $\hat\ell>0$ ($\hat\ell<0$) corresponds to increasing (decreasing) $\phi$.
For (ii) we use the $\theta$ equation in the conformal metric $\hat g_{\mu\nu}=\Omega^{2}g_{\mu\nu}$. At $\dot\theta=0$ (but with
$r$ arbitrary) the only $\theta$-dependence entering the Lagrangian is
through $\hat{g}_{\phi\phi}=P\,C^{2}r^{2}\sin^{2}\theta$, hence the
equation reduces to
\begin{equation}
\partial_{\theta}\,\bigl[P(\theta)\sin^{2}\theta\bigr]\Big|_{\theta=\theta_{\gamma}}=0.
\label{eq:EL_cone_condition}
\end{equation}
Equivalently, one may differentiate $\Theta(\theta)=P(\theta)-\hat\ell^{2}/\sin^{2}\theta$ and, using
$\Theta(\theta_{\gamma})=0 \Leftrightarrow \hat\ell^{2}=P(\theta_{\gamma})\sin^{2}\theta_{\gamma}$, impose $\Theta'(\theta_{\gamma})=0$; this yields exactly the same condition as Equation \eqref{eq:EL_cone_condition}. Writing $\nu=\cos\theta$ gives (for $0<a<\tfrac12$)
\begin{equation}
\frac{d}{d\nu}\,\bigl[(1+2a\nu)(1-\nu^{2})\bigr]=0
\;\,\Rightarrow\;\,
3a\,\nu_{\gamma}^{2}+\nu_{\gamma}-a=0.
\end{equation}
The physically admissible root (the only one with $|\nu_{\gamma}|\le1$) is
\begin{equation}\label{eq:cos_theta_gamma}
\cos\theta_{\gamma}
=\nu_{\gamma}
=\frac{-1+\sqrt{\,1+12a^{2}\,}}{6a} = \frac{2a}{1+\sqrt{1+12a^2}},
\end{equation}
where $a=\alpha m$ is dimensionless. Its small-$a$ expansion is
\begin{equation}
\cos\theta_{\gamma}
= a-3a^{3}+\mathcal O(a^{5})\ \ (a\ll1),
\label{eq:cone-angle}
\end{equation}
so $\theta_{\gamma}\to\frac{\pi}{2}$ as $a\to0$. Along the cone, Equation \eqref{eq:ellhat_cone} fixes the magnitude of
$\hat\ell$, and hence $d\phi/d\lambda_{K} =\hat\ell/\,\bigl[C\,P(\theta_{\gamma})\sin^{2}\theta_{\gamma}\bigr]$ is constant (its sign is the azimuthal orientation fixed by cone condition). Thus, except in the Schwarzschild limit, equatorial symmetry is broken: non-axial constant latitude null geodesics lie on the photon cone
$\theta=\theta_{\gamma}(a)$. The symmetry axes $(\theta=0,\pi)$ remain
excluded by the conical defect on one or both half-axes (see
section ~\ref{sec:metric}). To be more precise, along the cone the azimuthal constant is fixed in magnitude by Equation \eqref{eq:ellhat_cone}, thus
\begin{equation}\label{eq:ellhat}
    \hat\ell^{2}
= \frac{2\left(2+\sqrt{1+12a^{2}}\right)^{2}}
       {9\left(1+\sqrt{1+12a^{2}}\right)}
= 1+a^{2}-2a^{4}+\mathcal O(a^{6}) ,
\end{equation}
while the sign of $\hat\ell$ (equivalently, of $L$) sets the sense of
revolution via $d\phi/d\lambda_{K}
=\hat\ell/\bigl[C\,P(\theta_\gamma)\sin^{2}\theta_\gamma\bigr]=\text{const}$. In Figure \ref{Fig:theta_Lvsalpha} we illustrate the impact of the angular momentum on the type of latitudinal motion. 



\subsection{Photon surface (constant radius) and critical radius}
\label{sec:photon surface}
We repeat the radial first integral here for convenience
\begin{equation}
\Bigl(\frac{d\xi}{d\lambda_{K}}\Bigr)^{2}
= R(\xi):=\xi^{4}\hat e^{2}-\Delta(\xi),
\end{equation}
cf.\ Equations~\eqref{eq:first_r} and \eqref{eq:Delta_dimless}. A $\xi$-constant null orbit requires a double root of $R$:
\begin{equation}
R(\xi_{\rm ph})=0,\qquad R'(\xi_{\rm ph})=0.
\end{equation}
Eliminating $\hat e^{2}$ between the two conditions gives an equation independent of the constants of motion,
\begin{equation}
\frac{4\,\Delta(\xi)}{\xi}=\Delta'(\xi),
\quad\Rightarrow\quad
\xi_{\rm ph}(a)=\frac{6}{\,1+\sqrt{1+12a^{2}}\,}\  .
\label{eq:xi_ph}
\end{equation}
Restoring dimensions ($r=m\xi$, $a=\alpha m$), gives the (aspherical) photon surface radius inside the static patch
\begin{align*}
r_{\rm ph}(\alpha)=\frac{6m}{\,1+\sqrt{1+12\alpha^{2}m^{2}}\,},
\end{align*}
which expands as
\begin{align}
r_{\rm ph}
&=3m\Bigl[1-3(\alpha m)^{2}+18(\alpha m)^{4}+\mathcal O\,\bigl((\alpha m)^{6}\bigr)\Bigr].
\label{eq:rph}
\end{align}
 
\begin{figure}
\centering
\includegraphics[width=0.47\textwidth]{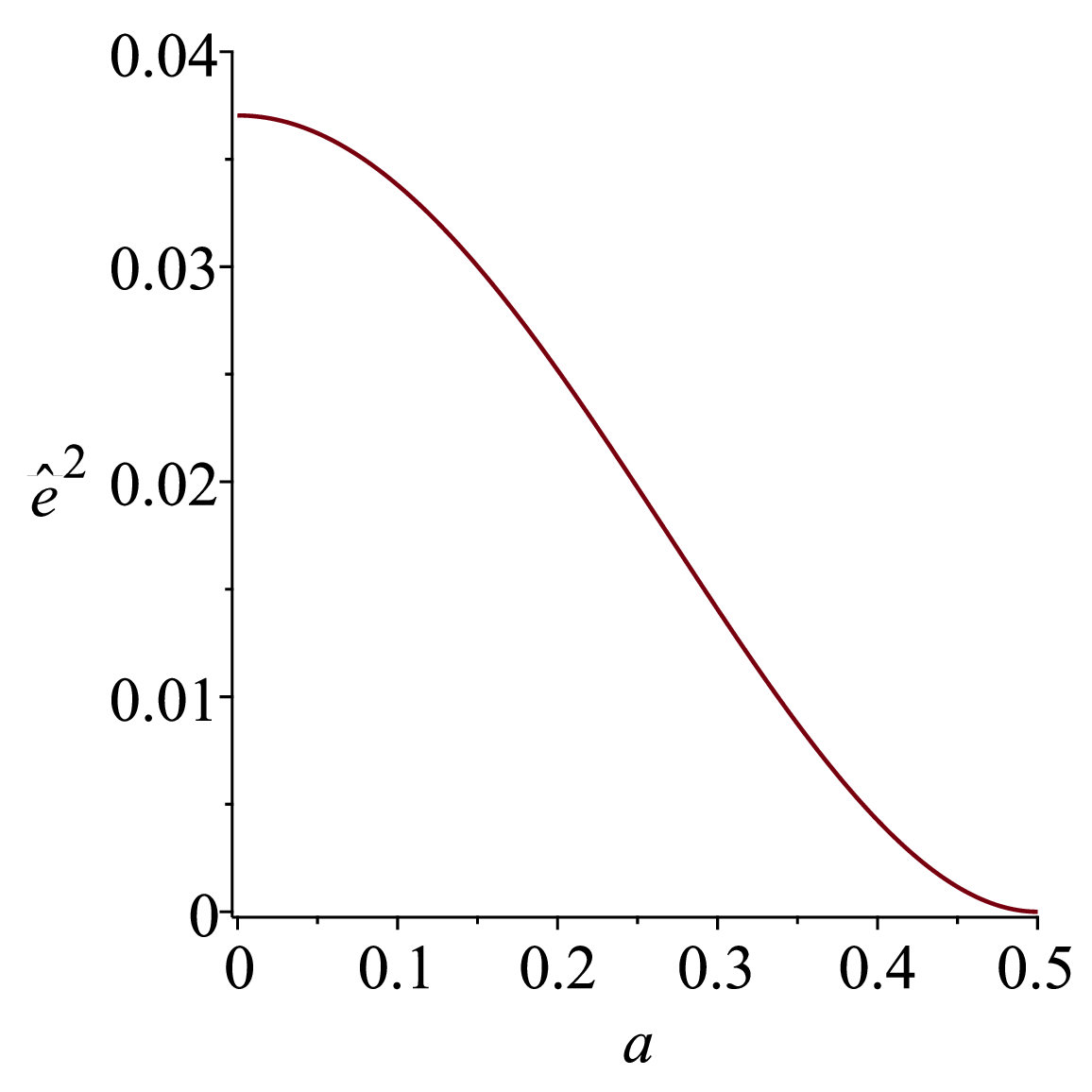}\quad
\includegraphics[width=0.47\textwidth]{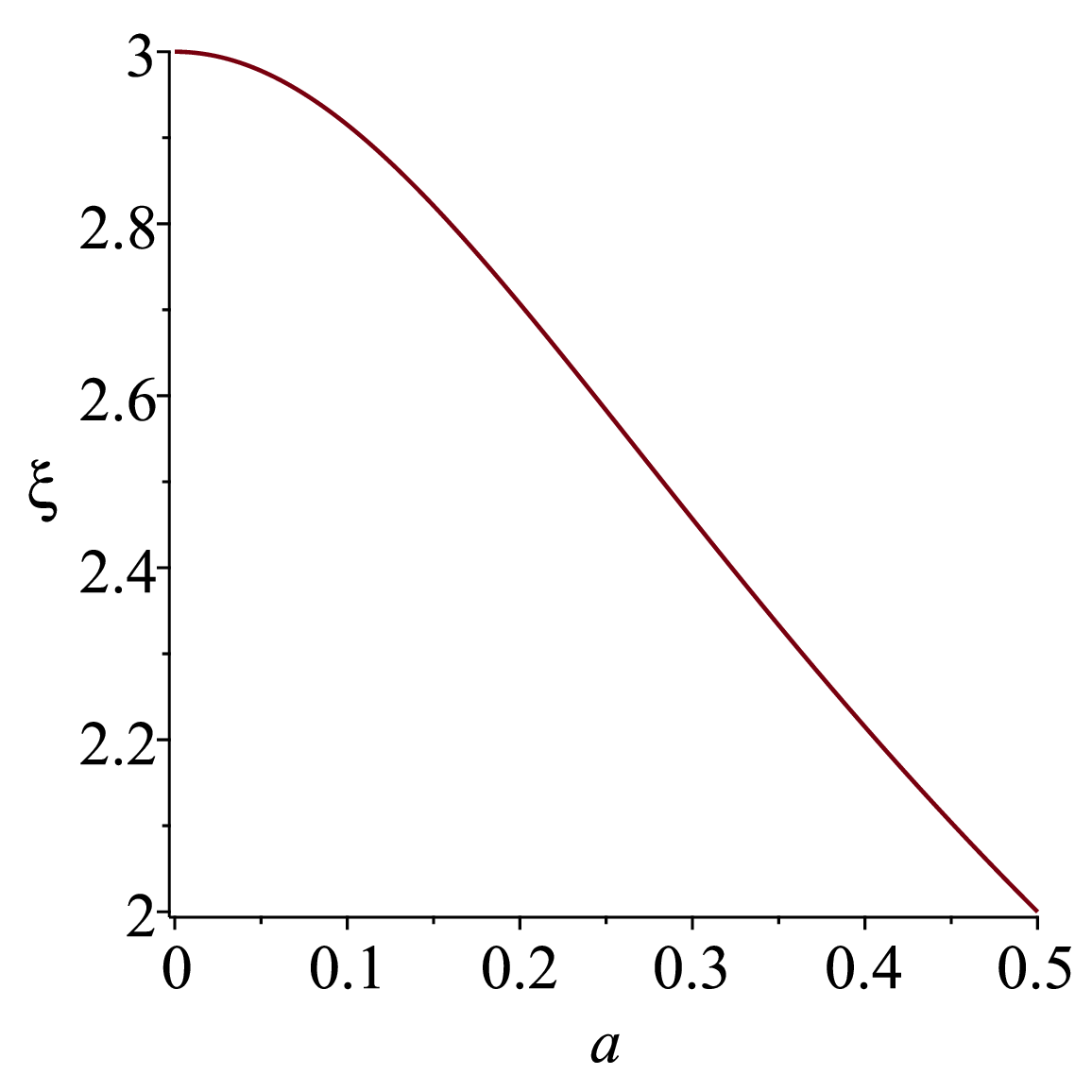}
\caption{Radial motion and photon surface.
Left: the critical ratio $\hat{e}^2:=\varepsilon^2/K=(\xi_{\rm ph}-2)^2/\xi_{\rm ph}^3$ as a function of $a$. Below the curve $R(\xi)=\varepsilon^{2}\xi^{4}-\Delta K$
has radial turning points; on the curve the orbit is spherical (double root); above it there are no turning points. 
Right: photon surface radius $\xi_{\rm ph}(a)=6/[1+\sqrt{1+12a^{2}}]$. \label{fig:radialmotion} }
\label{Fig:rmotion}
\end{figure}


\paragraph*{Energy parameter on the photon surface.} From $R(\xi_{\rm ph})=0$ we have

\begin{equation}
\hat e^{2} \;=\; \frac{\Delta(\xi_{\rm ph})}{\xi_{\rm ph}^{4}}\;.
\label{eq:ehat_from_R0}
\end{equation}
Using the photon surface condition \eqref{eq:xi_ph} one can show that on $\xi=\xi_{\rm ph}$ the factor $(1-a^{2}\xi^{2})$ equals
$\xi-2$, hence
\begin{equation}
\Delta(\xi_{\rm ph})
=\xi_{\rm ph}(\xi_{\rm ph}-2)^{2}.
\label{eq:Delta_identity_on_ph}
\end{equation}
Substituting Equation \eqref{eq:Delta_identity_on_ph} into
Equation \eqref{eq:ehat_from_R0} gives the compact expression
\begin{equation}
 \hat e^{2}
=\frac{(\xi_{\rm ph}-2)^{2}}{\xi_{\rm ph}^{3}}\,,
\label{eq:ehat_ph_final}
\end{equation}
Expanding for $a\ll1$ (so $\xi_{\rm ph}=3-9a^{2}+54a^{4}+\cdots$) yields

\begin{equation}\label{eq:ehat}
\hat e^{2}\;=\;\frac{1}{27}-\frac{a^{2}}{3}
+a^{4}+\mathcal O(a^{6}).
\end{equation}
as we see in Figure \ref{fig:radialmotion} (left). Note that, in unscaled constants this corresponds to
$K=\varepsilon^{2}\xi_{\rm ph}^{4}/\Delta(\xi_{\rm ph})$ and hence
$\hat e^{2}=\varepsilon^{2}/K$, in agreement with
the spherical orbit relations used elsewhere in the paper.

\subsection{Impact parameters}
\label{sec:impact-parameters}
Let $b:=L/E$ be the (axial) impact parameter ratio and
$\beta:=b/m$ its dimensionless counterpart.\footnote{In the spherically symmetric Schwarzschild case ($a=0$, $C=1$) the separation constant reduces to the total angular momentum squared, $K=L_{\rm tot}^{2}$, hence $\sqrt{Q}=\sqrt{K}/E=L_{\rm tot}/E$ is the usual (orientation-independent) impact parameter. Because one may rotate the orbital plane into the equator, one can identify $L_{\rm tot}=|L|$ and recover $b=L/E$. In the static C-metric the acceleration axis selects a preferred direction, so $L$ is genuinely the conserved axial component and the capture threshold depends on both $b=L/E$ and $Q=K/E^{2}$ through the factor $C^{2}P(\theta)\sin^{2}\theta$.}

\paragraph*{(i) Circular ring on the photon cone $(\xi=\xi_{\rm ph},\,\theta=\theta_\gamma)$.}

At fixed latitude the polar first integral enforces
\begin{equation}
K=\frac{\ell_C^{2}}{P(\theta)\,\sin^{2}\theta}\,,
\end{equation}
while the radial double root condition gives
\begin{equation}
K=\frac{\varepsilon^{2}\,\xi^{4}}{\Delta(\xi)}\,.
\end{equation}
Combining these at $(\xi,\theta)=(\xi_{\rm ph},\theta_\gamma)$ and restoring
dimensions yields
\begin{equation}
b_{\rm ring}^{2}(a,C)
= C^{2}\,P(\theta_{\gamma})\sin^{2}\theta_{\gamma}\,
   \frac{r_{\rm ph}^{4}}{\Delta(r_{\rm ph})}\; 
\label{eq:b_ring}
\end{equation}
where $P(\theta_{\gamma})=1+2a\cos\theta_{\gamma}$ and
$\cos\theta_{\gamma}$ is given by Equation~\eqref{eq:cos_theta_gamma} \footnote{Although our Weierstrass treatment avoids solving for turning points, one can recover the familiar turning-point relations as a consistency check: at any polar turning latitude $\theta_*$ (where $\dot\theta=0$) and any radial turning point $r_*$ (where $\dot r=0$) one has
$K=\ell_C^{2}/[P(\theta_*)\sin^{2}\theta_*]=\varepsilon^{2}\xi_*^{4}/\Delta(\xi_*)$,
hence $b^{2}=C^{2}P(\theta_*)\sin^{2}\theta_*\,r_*^{4}/\Delta(r_*)$.
For the spherical cone orbit $(r_*,\theta_*)=(r_{\rm ph},\theta_\gamma)$ this reduces to Equation \eqref{eq:b_ring}.}.
For $a\ll1$, one has (using $\cos\theta_\gamma=a-3a^{3}+\mathcal O(a^{5})$)

\begin{equation}
P(\theta_{\gamma})\sin^{2}\theta_{\gamma}
=1+a^{2}-2a^{4}+\mathcal O(a^{6}),
\end{equation}
and using Equation \eqref{eq:rph} for $a\ll1$ one finds

\begin{equation}\label{eq:bring}
b_{\rm ring}
=3\sqrt{3}\,m\,C\Bigl[1+5a^{2}+\mathcal O(a^{4})\Bigr].
\end{equation}
Or equivalently, substituting the exact relations $\xi_{\rm ph}(a)$ and $\theta_\gamma(a)$ from
Equations~\eqref{eq:xi_ph} and \eqref{eq:cos_theta_gamma} into Equation \eqref{eq:b_ring} yields the closed form

\begin{equation}
\frac{b_{\rm ring}}{mC}
= \frac{2\sqrt{3}\,\bigl(2+\sqrt{1+12a^{2}}\bigr)}
        {\bigl(1+\sqrt{1+12a^{2}}\bigr)\,\bigl(2-\sqrt{1+12a^{2}}\bigr)} ,
\end{equation}
which expands also as $b_{\rm ring}=3\sqrt{3}\,m\,C\,[\,1+5a^{2}+\mathcal O(a^{4})\,]$ .

The $C$-dependence enters only through the azimuthal constant (via $g_{\phi\phi}\propto C^{2}$); for a fixed cone latitude
the true capture/escape threshold is $b_{\rm ring}$, not a radius-only proxy.

It is sometimes convenient to factor out the inclination/conicity and quote the
radius-only scale impact parameter $b_{\rm rad}$, which equals the true threshold in spherical symmetry but serves only as a baseline in the anisotropic C-metric:

\paragraph*{(ii) Radius-only proxy.}

The radius-only impact parameter scale is given by
\begin{equation}
b_{\rm rad}(r_{\rm ph})
:=\frac{r_{\rm ph}^{2}}{\sqrt{\Delta(r_{\rm ph})}},
\label{eq:b_proxy}
\end{equation}
which is obtained by ignoring the latitude/conicity factor in the angular barrier. In spherically symmetric spacetimes (e.g., Schwarzschild, where $P\equiv1$ and $C\equiv1$) this coincides with the true critical impact parameter for photon capture, because the spherical photon orbit fixes a single radius $r_{\rm ph}$ and the effective potential depends only on $r$. In the static C-metric the photon surface still sits at a single radius
$r_{\rm ph}(\alpha)$ [Equation \eqref{eq:xi_ph}], but the angular sector is
anisotropic through $P(\theta)=1+2a\cos\theta$ and the conical parameter
$C$. As a result, the true threshold depends on the latitude of
motion and reads (our Equation \eqref{eq:b_ring})
\begin{align}
b_{\rm ring}^{2}
&= C^{2}\,P(\theta_{\gamma})\sin^{2}\theta_{\gamma}\,
\frac{r_{\rm ph}^{4}}{\Delta(r_{\rm ph})}\nonumber\\
&=\underbrace{C^{2}P(\theta_{\gamma})\sin^{2}\theta_{\gamma}}_{\text{inclination/conicity}} \;\underbrace{b_{\rm rad}^{2}}_{\text{radius-only scale}}.
\end{align}
showing explicitly how orientation ($\theta_\gamma$) and conicity ($C$)
enter only through $b_{\rm ring}$, whereas $b_{\rm rad}$ captures the
radius-only part common to all latitudes \footnote{Since $b_{\rm ring}=b_{\rm rad}\,\sqrt{P(\theta_\gamma)\sin^{2}\theta_\gamma}\,C$,
their small‑$a$ series are consistent:
$\sqrt{P(\theta_\gamma)\sin^{2}\theta_\gamma}
=1+\tfrac12 a^{2}+\mathcal O(a^{4})$, hence the net $+5a^{2}$ coefficient
in $b_{\rm ring}$ follows from
$\tfrac{9}{2}+\tfrac12=5$.}. Equivalently, in dimensionless form with $\beta:=b/m$, one has
\begin{align}
&\beta_{\rm rad}^{2}=\frac{b_{\rm rad}^{2}}{m^{2}}
= \frac{\xi_{\rm ph}^{4}}{\Delta(\xi_{\rm ph})}
=:Q_{\rm ph}(a),\nonumber \\
&\beta_{\rm ring}^{2}= C^{2}P(\theta_\gamma)\sin^{2}\theta_\gamma\,Q_{\rm ph}(a),
\end{align}
therefore, $Q_{\rm ph}(a)$ is the orientation-independent baseline set by
the spherical photon surface, while the factor
$C^{2}P(\theta_{\gamma})\sin^{2}\theta_{\gamma}$ encodes inclination and
conical deficit. Thus:
\begin{itemize}\setlength\itemsep{2pt}
\item In Schwarzschild ($a=0$, $C=1$) one recovers $b_{\rm ring}=b_{\rm rad}=3\sqrt{3}\,m$.
\item In the static C-metric, $b_{\rm rad}$ remains a useful reference
scale that isolates the radial physics of the photon surface, but it
does not by itself decide capture/escape; thus the 
latitude-aware threshold is $b_{\rm ring}$.
\end{itemize}
For small $a$, by using Equations \eqref{eq:xi_ph}-\eqref{eq:rph} one finds
\begin{equation}
b_{\rm rad}
=3\sqrt{3}\,m\Bigl[1+\tfrac{9}{2}a^{2}+\mathcal O(a^{4})\Bigr].
\end{equation}
Note that: (i) Constant latitude null geodesics at the cone angle
Equation~\eqref{eq:cos_theta_gamma} are elementary in Hong-Teo coordinates (fixed $x=\cos\theta$). (ii) The constant radius condition \eqref{eq:xi_ph} matches optical metric arguments for photon surfaces in accelerated black holes; see, e.g., \cite{2016PhLB..763..169G,PhysRevD.103.024007}
for complementary perspectives.\footnote{In Hong-Teo coordinates $(x,y)$ one has $x=\cos\theta$ and, in the Ricci-flat case, $r=-(\alpha y)^{-1}$. Fixed $x$ null orbits arise for specific ratios of the conserved quantities; see \cite{PhysRevD.103.024007}.}


%


\section{Astrophysical implications}
\label{sec:astro}

We specialise to a static observer at $(\xi_{o},\theta_{o})$
inside the static patch $2<\xi_{o}<1/a$ (with $a=\alpha m\in(0,\tfrac12)$).
An orthonormal tetrad aligned with the coordinate axes is
\begin{subequations}
\begin{align}
e_{(t)}&=\frac{\Omega_{o}\,\xi_{o}}{\sqrt{\Delta_{o}}}\,\partial_{t}, \\
e_{(r)}&=\frac{\Omega_{o}\,\sqrt{\Delta_{o}}}{\xi_{o}}\,\partial_{r}, \\
e_{(\theta)}&=\frac{\Omega_{o}\,\sqrt{P_{o}}}{\xi_{o}}\,\partial_{\theta},\\
e_{(\phi)}&=\frac{\Omega_{o}}{C\,\xi_{o}\sin\theta_{o}\sqrt{P_{o}}}\,\partial_{\phi},
\label{eq:tetrad_obs}
\end{align}    
\end{subequations}

with $\Omega=1+a\xi\cos\theta$, $P=1+2a\cos\theta$, and
$\Delta=\xi^{2}(1-\tfrac{2}{\xi})(1-a^{2}\xi^{2})$. A subscript $o$
means “evaluate at the observer”. One can check that
$g(e_{(t)},e_{(t)})=-1$ and $g(e_{(i)},e_{(j)})=\delta_{ij}$.

For a photon with conserved $(E,L)$ and separation constant $K$
(section \ref{sec:null-dimless}), we define the screen coordinates
$(\iota,\vartheta)$ in the observer’s sky by 
\begin{equation}
\iota:=-\frac{k_{(\phi)}}{k_{(t)}},\qquad
\vartheta:=+\frac{k_{(\theta)}}{k_{(t)}},\qquad
k_{(a)}:=g_{\mu\nu}\,k^{\mu}e_{(a)}^{\nu}.
\end{equation}
Using Equation \eqref{eq:tetrad_obs} together with the first integrals in
Equation \eqref{eq:firstorder_dimless} gives
\begin{subequations}
\begin{align}
\iota&= \frac{\sqrt{\Delta_{o}}}{C\,\xi_{o}^{2}\sin\theta_{o}\sqrt{P_{o}}}\;
    \frac{L}{E},\label{eq:alpha_general}\\
\vartheta
&= \frac{\sqrt{\Delta_{o}}}{\xi_{o}^{2}}\;
    \sqrt{\,Q-\frac{1}{C^{2}P_{o}\sin^{2}\theta_{o}}
 \Bigl(\frac{L}{E}\Bigr)^{2}},
\label{eq:beta_general}
\end{align}    
\end{subequations}
with $Q:=K/E^{2}$ as in the section \ref{sec:null-dimless}.
(Our sign choice makes $\iota$ increase toward $+e_{(\phi)}$ and
$\vartheta$ toward $+e_{(\theta)}$.) We see here that the conicity parameter enters both $\iota$ and $\vartheta$ via $L/(CE)$. Summing the squares of Equations \eqref{eq:alpha_general} and \eqref{eq:beta_general} gives the useful conicity-independent identity
\begin{equation}
\iota^{2}+\vartheta^{2}
=\frac{\Delta_{o}}{\xi_{o}^{4}}\;Q.
\label{eq:alpha2plusbeta2}
\end{equation}
This implies that a change in the conicity $C$ induces a rotation of the point $(\iota,\vartheta)$ on the screen.

\subsection{Shadow boundary from the photon surface}
\label{subsec:shadow}

From Equation ~\eqref{eq:xi_ph}, the (spherical) photon surface sits at the
constant radius
\begin{equation}\label{eq:xi}
\xi_{\rm ph}(a)=\frac{6}{1+\sqrt{1+12a^{2}}}.
\end{equation}
For any spherical null orbit one has (section \ref{sec:null-dimless})
\begin{equation}
Q=\frac{\xi_{\rm ph}^{4}}{\Delta(\xi_{\rm ph})}
=:Q_{\rm ph}(a),
\label{eq:Q_spherical}
\end{equation}
which is independent of $L/E$. Every generator of the shadow boundary arises from
such an orbit, hence satisfies Equation \eqref{eq:Q_spherical}. Eliminating $L/E$
between Equation \eqref{eq:alpha_general}-Equation \eqref{eq:beta_general} using Equation 
\eqref{eq:Q_spherical} collapses the locus to
\begin{equation}
\iota^{2}+\vartheta^{2}
= \frac{\Delta(\xi_{o})}{\xi_{o}^{4}}\;
  \frac{\xi_{\rm ph}(a)^{4}}
       {\Delta\,\bigl(\xi_{\rm ph}(a)\bigr)} .
\label{eq:shadow_circle}
\end{equation}
For any static observer in the static C-metric, the black hole shadow
boundary is a circle in the local sky. As local Lorentz transformations act conformally on the celestial sphere and map circles to circles, this also holds for any observer. This generalises the result in \cite{2015IJMPD..2442024G} on the circular shadow boundary to the case that the conicity parameter $C$ is explicitly included into the equations. As we see in Equation \eqref{eq:alpha_general} and Equation \eqref{eq:beta_general}, the conicity enters into the cartesian-like coordinates on the observer screen for any single null geodesic, but drops out of the radius of the shadow due to its circularity. The (screen) radius \footnote{It is $R_{\rm sh} = \sin\theta \approx \theta$, where $2\theta$ is the angular diameter of the shadow.} is
\begin{equation}
R_{\rm sh}(\xi_{o},\theta_{o};a)
= \frac{\sqrt{\Delta(\xi_{o})}}{\xi_{o}^{2}}\;
  \frac{\xi_{\rm ph}(a)^{2}}
{\sqrt{\Delta\,\bigl(\xi_{\rm ph}(a)\bigr)}} ,
\label{eq:Rshadow}
\end{equation}
with $\xi_{\rm ph}(a)$ in Equation \eqref{eq:xi}. Note that it does not depend on the inclination of the observer $\theta_o$. Moreover, neither the centre nor the radius depends on the conicity $C$. Indeed, as $a\to0$ one recovers the Schwarzschild result
\begin{equation}
R_{\rm sh}
=3\sqrt{3}\,\sqrt{1-\tfrac{2}{\xi_{o}}}\,/\xi_{o},    
\end{equation}
corresponding to the usual $3\sqrt{3}\,m$ scaled by the redshift to the
observer at $\xi_{o}$.

Using Equation \eqref{eq:Rshadow} and the photon surface identity
$\Delta(\xi_{\rm ph})=\xi_{\rm ph}(\xi_{\rm ph}-2)^{2}$ (which follows immediately from the double root condition
$4\Delta/\xi=\Delta'$ at $\xi=\xi_{\rm ph}$), the shadow radius can be
factorised as
\begin{equation}
R_{\rm sh}(\xi_o,\theta_o;a)
=\frac{\sqrt{(1-2/\xi_o)\,\bigl(1-a^{2}\xi_o^{2}\bigr)}}{\xi_o}\;
  \frac{\xi_{\rm ph}(a)^{3/2}}{\xi_{\rm ph}(a)-2},
\label{eq:Rsh_factorised}
\end{equation}
where $\xi_{\rm ph}$ is given by Equation \eqref{eq:xi} (or Equation~\eqref{eq:xi_ph}). 
 
This makes the competing effects of the acceleration explicit: a local redshift factor
$\sqrt{1-a^{2}\xi_o^{2}}$ that decreases with $a$ (and
vanishes as the acceleration horizon $\xi_{A}=1/a$ approaches
$\xi_o$), and a photon–surface factor
$\xi_{\rm ph}^{3/2}/(\xi_{\rm ph}-2)$ that increases as
$\xi_{\rm ph}\downarrow2$ when $a\uparrow 1/2$.

A direct differentiation gives the clean, fully reduced identity
\begin{equation}
\frac{d}{da}\ln R_{\rm sh}
=-\frac{a\,\xi_o^{2}}{1-a^{2}\,\xi_o^{2}}
+\frac{a\,\xi_{\rm ph}(a)^{2}}{\xi_{\rm ph}(a)-2}.
\label{eq:dlogR_da_clean}
\end{equation}
Equivalently, using $1+2a^{2}\xi_{\rm ph}=(6-\xi_{\rm ph})/\xi_{\rm ph}$, which follows from the photon–surface relation $a^{2}=(3-\xi_{\rm ph})/\xi_{\rm ph}^{2}$, one may write
\begin{equation*}
\frac{d}{da}\ln R_{\rm sh}
=-\frac{a\,\xi_o^{2}}{1-a^{2}\,\xi_o^{2}}
+\frac{a\,\xi_{\rm ph}(a)\,\bigl[6-\xi_{\rm ph}(a)\bigr]}
       {\bigl[\xi_{\rm ph}(a)-2\bigr]\,
        \bigl[1+2a^{2}\,\xi_{\rm ph}(a)\bigr]}.
\label{eq:dlogR_da_equiv}
\end{equation*}
In this form, one sees that the photon–surface contribution is positive for $2<\xi_{\rm ph}<3$, while the local term is always negative on $0<a<1/\xi_o$. Their competition yields the following:

\begin{itemize}
\item Small-$a$ behaviour. Expanding Equation \eqref{eq:Rshadow} and
Equation \eqref{eq:xi} one finds
\begin{equation}
\frac{R_{\rm sh}(a)}{R_{\rm sh}(0)}
=1+\frac{1}{2}\,(9-\xi_o^{2})\,a^{2}
+\mathcal{O}(a^{4}).
\label{eq:Rsh_small_a}
\end{equation}
Hence $R_{\rm sh}$ initially increases with $a$ if
$2<\xi_o<3$, is flat at $\xi_o=3$, and decreases if
$\xi_o>3$.

\item Global trend on the allowed range $0<a<1/\xi_o$.
Because the prefactor $\sqrt{1-a^{2}\xi_o^{2}}$ in Equation 
\eqref{eq:Rsh_factorised} vanishes as $a\to(1/\xi_o)^{-}$, one always
has $R_{\rm sh}(a)\to0$ as the observer approaches the acceleration
horizon. Consequently, for $\xi_o>3$ the shadow radius decreases
monotonically with $a$. For $2<\xi_o<3$, $R_{\rm sh}(a)$ increases
at small $a$ (by Equation \eqref{eq:Rsh_small_a}), reaches a single maximum at
some $a_{*}(\xi_o)\in(0,1/\xi_o)$, and then decreases to zero as
$a\to(1/\xi_o)^{-}$.
\end{itemize}

In all cases the centre of the shadow remains unshifted and the boundary
remains exactly circular; only its angular scale varies with $a$ and the observer position.

In summary, acceleration ($a>0$) changes the "size" but not the "shape" (still a circle) and does not shift the centre. Although the C-metric breaks equatorial symmetry in the photon dynamics (cone angle, polar turning points), the entire family of spherical photon orbits sits at a single radius $\xi_{\rm ph}(a)$. Inserting the corresponding $Q_{\rm ph}(a)$ into Equation \eqref{eq:alpha2plusbeta2} yields the circular boundary \eqref{eq:shadow_circle}. 

Importantly, all spherical photon orbits share a single radius in the static C-metric. More precisely, in the $K$-scaled Mino system the radial equation reads
\begin{equation}
\Bigl(\tfrac{d\xi}{d\lambda_{K}}\Bigr)^{2}
= R(\xi):=\xi^{4}\hat e^{2}-\Delta(\xi),
\qquad
\hat e:=\varepsilon/\sqrt{K},
\label{eq:R_spherical_expl}
\end{equation}
cf.\ Equation \eqref{eq:first_r}. A constant radius "spherical" null orbit requires a double root of $R$:
\begin{equation}
R(\xi_{\rm ph})=0,
\qquad
R'(\xi_{\rm ph})=0.
\end{equation}
From $R(\xi_{\rm ph})=0$ one finds $\hat e^{2}=\Delta(\xi_{\rm ph})/\xi_{\rm ph}^{4}$. Inserting this into
$R'(\xi)=4\xi^{3}\hat e^{2}-\Delta'(\xi)$ gives
\begin{equation}
\frac{4\,\Delta(\xi_{\rm ph})}{\xi_{\rm ph}}=\Delta'(\xi_{\rm ph}),
\end{equation}
which is exactly the photon surface condition \eqref{eq:xi_ph}. Thus $\xi_{\rm ph}=\xi_{\rm ph}(a)$ is uniquely fixed by the metric function $\Delta$ and depends only on the acceleration parameter $a$.

Structurally, this uniqueness follows because in the conformal metric $\hat g_{\mu\nu}=\Omega^{2}g_{\mu\nu}$ the radial potential Equation \eqref{eq:R_spherical_expl} depends on a single combination of constants of motion ($\hat e^{2}$) and on $\xi$ alone; the two double root conditions then determine $(\xi_{\rm ph},\hat e^{2})$ with no residual dependence on the polar constant. This contrasts with Kerr, where the radial potential depends on two independent ratios (e.g.\ $L_{z}/E$ and $Q/E^{2}$), so the double–root system solves for those ratios as functions of the radius, leaving a one-parameter family of spherical photon radii. Here, by static assumption and the resulting radial decoupling, all spherical photon orbits lie on the single photon surface $\xi=\xi_{\rm ph}(a)$.

It is worth mentioning here that the stationary, axisymmetric barotropic thick accretion discs  in this set-up exist only for relatively small $\alpha$ and vanish as $\alpha$ increases \cite{2022PhRvD.105j3017F,2023mgm..conf..317F}. The disc affects the brightness map (which photons are emitted or absorbed), not the circular capture boundary itself. For typical observers with $\xi_o>3$, $R_{\rm sh}$ decreases monotonically as $a$ increases, while the band of stable circular geodesics (and hence the geometric support for a thick torus) also shrinks and disappears
at $a_{\rm crit}$. Thus the two findings are consistent: larger acceleration simultaneously reduces the observable shadow scale and suppresses stationary thick discs. However, to quantitatively relate disc morphology to observed images one must fix
an emission model (optically thin flow vs.\ optically thick torus), but the geometric inputs are already provided here. A minimal, self-consistent pipeline is: (i) choose a stationary torus family (e.g.\ constant $\ell_C$); (ii) locate the centre and cusp and the Hessian of effective potential \cite{2022PhRvD.105j3017F}; (iii) ray traced photons using the closed null solution (section \ref{sec:null-dimless}) and the observer tetrad Equation \eqref{eq:tetrad_obs}. The resulting images will always contain the circular shadow boundary Equation \eqref{eq:shadow_circle}; the disc merely modulates the brightness across it.

While the circular capture boundary itself is $C$-independent, the lensed image of an extended emitter (such as a disc/torus) can depend on $C$ through the global lensing map; see Section~\ref{subsec:string_bound}.



\subsection{Inferring the acceleration from a single shadow}
\label{subsec:inference}

From Equations \eqref{eq:shadow_circle}-\eqref{eq:Rshadow}, a measured screen radius $R_{\rm sh}$ at a known mass-distance relation $\xi_o$ determines the dimensionless combination
\begin{equation}
\mathcal{A}:=\frac{\xi_o^{2}}{1-2/\xi_o}\,R_{\rm sh}^{2}.
\label{eq:inference_A}
\end{equation}
Using $\Delta(\xi_o)=\xi_o^{2}(1-2/\xi_o)(1-a^{2}\xi_o^{2})$ and the photon surface identity
$\Delta(\xi_{\rm ph})=\xi_{\rm ph}(\xi_{\rm ph}-2)^{2}$, Equation \eqref{eq:Rshadow} gives
\begin{equation}
\mathcal{A}\,(\xi_{\rm ph}-2)^{2}
=\bigl(1-a^{2}\xi_o^{2}\bigr)\,\xi_{\rm ph}^{3}.
\label{eq:inference_A_relation}
\end{equation}
Eliminating $a^{2}$ via the photon surface relation
$a^{2}=(3-\xi_{\rm ph})/\xi_{\rm ph}^{2}$ yields the cubic equation
\begin{equation}
\mathcal{A}\,(\xi_{\rm ph}-2)^{2}
=\xi_{\rm ph}^{3}+\xi_o^{2}\xi_{\rm ph}^{2}-3\xi_o^{2}\xi_{\rm ph},
\label{eq:inference_cubic_xiph}
\end{equation}
equivalently,
\begin{equation}
\xi_{\rm ph}^{3}+(\xi_o^{2}-\mathcal{A})\,\xi_{\rm ph}^{2}
+\bigl(4\mathcal{A}-3\xi_o^{2}\bigr)\,\xi_{\rm ph}-4\mathcal{A}=0.
\label{eq:inference_cubic_poly}
\end{equation}
Given $\mathcal{A}$ from Equation \eqref{eq:inference_A}, one solves Equation \eqref{eq:inference_cubic_poly} and selects the physical root in the static range $2<\xi_{\rm ph}<3$.

Finally, the dimensionless acceleration parameter $a=\alpha m$ follows algebraically from the photon surface relation (equivalent to the double root condition)
\begin{equation}
a^{2}=\frac{3-\xi_{\rm ph}}{\xi_{\rm ph}^{2}},
\qquad
0<a<\tfrac12\ \Longleftrightarrow\ 2<\xi_{\rm ph}<3.
\label{eq:inference_a}
\end{equation}
Therefore, given a measured screen radius $R_{\rm sh}$ and a known observer position $\xi_o=r_o/m$, one first forms the dimensionless combination $\mathcal{A}$ from Equation~\eqref{eq:inference_A}. The photon surface radius $\xi_{\rm ph}$ is then obtained by solving the cubic Equation~\eqref{eq:inference_cubic_poly} and selecting the physical root in the static range $2<\xi_{\rm ph}<3$. Finally, the dimensionless acceleration $a=\alpha m$ follows algebraically from Equation~\eqref{eq:inference_a}.

Without an independent determination of $\xi_o$, the shadow radius $R_{\rm sh}$ alone leaves a degeneracy between $a$ and $\xi_o$. Furthermore, error propagation is straightforward: since $\mathcal{A}\propto R_{\rm sh}^{2}$ one has
$\delta\mathcal{A}/\mathcal{A}=2\,\delta R_{\rm sh}/R_{\rm sh}$; then
$\delta\xi_{\rm ph}$ follows from implicit differentiation of
Equation~\eqref{eq:inference_cubic_poly}, and $\delta a$ from
$\partial a/\partial\xi_{\rm ph}$ using Equation~\eqref{eq:inference_a}.



\subsection{Observer dependence of the shadow radius}
\label{subsec:shadow_observer_dependence}

The shadow radius derived in section~\ref{subsec:shadow} is a local
observable defined in the instantaneous rest frame of an observer. For a static observer at $(\xi_o,\theta_o)$ it is
\begin{equation}
R_{\rm sh}(\xi_o;a)
=\frac{\sqrt{(1-2/\xi_o)\,(1-a^{2}\xi_o^{2})}}{\xi_o}\;
  \frac{\xi_{\rm ph}(a)^{3/2}}{\xi_{\rm ph}(a)-2},
\label{eq:Rsh_static_recall}
\end{equation}
cf.\ Equation~\eqref{eq:Rsh_factorised}. Note that $R_{\rm sh}$ is
independent of $\theta_o$ and of the conicity parameter $C$.

\paragraph*{(i) Varying the observer position at fixed acceleration.}
For fixed $a\in(0,\tfrac12)$ the $\xi_o$-dependence of the relation
\eqref{eq:Rsh_static_recall} is entirely carried by the factor
\begin{equation}
\mathcal{R}(\xi_o;a)
:=\frac{\sqrt{(1-2/\xi_o)\,(1-a^{2}\xi_o^{2})}}{\xi_o},
\qquad 2<\xi_o<\frac{1}{a}.
\end{equation}
Equivalently,
\begin{equation}
\mathcal{R}(\xi_o;a)^{2}
=\frac{(\xi_o-2)\,(1-a^{2}\xi_o^{2})}{\xi_o^{3}}.
\label{eq:Rsh_xidep_sq}
\end{equation}
This vanishes at both Killing horizons, $\xi_o\downarrow2$ (black hole horizon) and $\xi_o\uparrow 1/a$ (acceleration horizon), and therefore has at least one maximum in the static patch. Differentiating
Equation \eqref{eq:Rsh_xidep_sq} yields
\begin{equation}
\frac{d}{d\xi_o}\,\mathcal{R}(\xi_o;a)^{2}
=-\,\frac{2}{\xi_o^{4}}\Bigl(a^{2}\xi_o^{2}+\xi_o-3\Bigr),
\end{equation}
so the maximum is unique and occurs at
\begin{equation}
a^{2}\xi_*^{2}+\xi_* - 3=0
\quad\Longrightarrow\quad
\xi_*(a)=\frac{-1+\sqrt{1+12a^{2}}}{2a^{2}}.
\label{eq:xi_star}
\end{equation}
Using Equation~\eqref{eq:xi_ph} one checks that this is exactly the
photon surface radius:
\begin{equation}
\xi_*(a)=\xi_{\rm ph}(a).
\end{equation}
Consequently, a static observer placed at $\xi_o=\xi_{\rm ph}(a)$ sees the shadow as a hemisphere:
\begin{equation}
R_{\rm sh}\bigl(\xi_o=\xi_{\rm ph}(a);a\bigr)=1
\quad(\text{i.e.\ }\Theta_{\rm sh}=\tfrac{\pi}{2}\text{ with }R_{\rm sh}=\sin\Theta_{\rm sh}).
\end{equation}
Thus, for fixed $a$, the shadow grows from $0$ at $\xi_o=2$ to its maximal value $R_{\rm sh}=1$ at $\xi_o=\xi_{\rm ph}(a)$, and then shrinks back to $0$ as the observer approaches the acceleration horizon.


\paragraph*{(ii) Varying the acceleration at fixed observer position.}
For fixed $\xi_o$, the behaviour of $R_{\rm sh}(\xi_o;a)$ as a function of $a$ is controlled by the competition already made explicit in Equation~\eqref{eq:Rsh_factorised}: the local factor
\begin{equation*}
 \sqrt{1-a^{2}\xi_o^{2}},   
\end{equation*}
decreases with $a$, while the photon surface factor 
\begin{equation*}
\xi_{\rm ph}^{3/2}/(\xi_{\rm ph}-2)    
\end{equation*}
increases as $a\uparrow\tfrac12$. As shown in section~\ref{subsec:shadow}, this implies that $R_{\rm sh}(a)$ is monotonically decreasing for $\xi_o>3$, while for $2<\xi_o<3$ it increases at small $a$ and then decreases to $0$ as $a\to(1/\xi_o)^{-}$.

\paragraph*{(iii) Moving observers: aberration at fixed event.}
All screen variables $(\iota,\vartheta)$ are defined in an orthonormal tetrad, so changing the observer state of motion at the same spacetime event is implemented by a local Lorentz transformation. For a purely radial motion (relative to the static tetrad) the shadow boundary remains a circle; only its angular radius changes by special relativistic effects.

Let $v \in(-1,1)$ be the signed radial 3-velocity measured in the static
tetrad, with $v>0$ meaning motion along $+e_{(r)}$ (outward, i.e.\ away from the black hole), and let $\gamma=(1-v^{2})^{-1/2}$. Writing the static-observer shadow half opening angle as $\Theta_{\rm sh}\in(0,\pi/2]$ with
\begin{equation}
R_{\rm sh}=\sin\Theta_{\rm sh},
\qquad
\cos\Theta_{\rm sh}=\sqrt{1-R_{\rm sh}^{2}},
\end{equation}
the aberrated radius seen by the moving observer is
\begin{equation}
R_{\rm sh}'
=\frac{R_{\rm sh}}{\gamma\Bigl(1+v\,\sqrt{1-R_{\rm sh}^{2}}\Bigr)}.
\label{eq:Rsh_aberration_radial}
\end{equation}
Hence outward motion ($v>0$) shrinks the shadow, while inward motion ($v<0$) enlarges it. For small speeds,
\begin{equation}
\frac{\delta R_{\rm sh}}{R_{\rm sh}}
\simeq -\,v\,\sqrt{1-R_{\rm sh}^{2}}
\qquad(|v|\ll1),
\end{equation}
with the sign convention stated above. For a nonradial boost the boundary is still mapped to a circle (aberration is conformal on the celestial sphere), but its centre is generally displaced; the radial case Equation \eqref{eq:Rsh_aberration_radial}
is the axisymmetric specialisation relevant when the motion is toward/away from the black hole.

\paragraph*{Small-$a$ scaling (acceleration vs.\ aberration).}
For a fixed observation event, the static-observer shadow radius
$R_{\rm sh}(\xi_o;a)$ is an even function of the acceleration parameter $a$: both $\Delta(\xi)$ and the photon surface radius $\xi_{\rm ph}(a)$ depend on $a$
only through $a^{2}$. Consequently, the leading acceleration-induced change of the shadow size is quadratic in $a$, as seen explicitly in the small-$a$ expansion Equation \eqref{eq:Rsh_small_a}. For a moving observer, aberration maps the local circular boundary to another circle (and generally shifts its centre for nonradial boosts), but the dependence on the spacetime parameters still enters
only through the local radius $R_{\rm sh}$ at the observation event. Therefore, the boosted radius inherits the same behaviour: at small $a$ the acceleration imprint remains $\mathcal{O}(a^{2})$, whereas the kinematic change from aberration is $\mathcal{O}(v)$ for small 3-velocity $v$. In practice, this means that for mildly accelerated black holes a modest peculiar velocity can dominate the apparent change in shadow size unless the observer's motion is independently constrained.



\subsection{Eikonal ringdown}
\label{sec:ringdown}

As shown in Equation~\eqref{eq:xi}, unstable spherical photon orbits lie on the photon surface
$\xi=\xi_{\rm ph}(a)$.
The fixed-latitude subset lies on the photon cone, with
\begin{equation}
\cos\theta_{\gamma}= \frac{-1+\sqrt{\,1+12a^{2}\,}}{6a},
\end{equation}
as derived in sections~\ref{sec:photon surface} and \ref{sec:photon-cone}.
We use $\Delta(\xi)=\xi^{2}\bigl(1-\tfrac{2}{\xi}\bigr)\bigl(1-a^{2}\xi^{2}\bigr)$
and $P(\theta)=1+2a\cos\theta$ throughout.

\subsubsection{Orbital angular velocity}

In the $K$-scaled Mino system (Section ~\ref{sec:null-dimless}),
\begin{equation*}
\frac{d\phi}{d\lambda_{K}}=\frac{\hat\ell}{CP(\theta)\sin^{2}\theta},\qquad
\frac{d\tau}{d\lambda_{K}}=\frac{\xi^{4}\hat e}{\Delta(\xi)} .
\end{equation*}
On a spherical cone orbit $(\xi,\theta)=(\xi_{\rm ph},\theta_{\gamma})$ the
constants obey
\begin{align}
K&=\frac{\varepsilon^{2}\xi_{\rm ph}^{4}}{\Delta(\xi_{\rm ph})}
=\frac{\ell_{C}^{2}}{P(\theta_{\gamma})\sin^{2}\theta_{\gamma}}\nonumber\\
&\Longrightarrow\quad
\hat e^{2}=\frac{\Delta(\xi_{\rm ph})}{\xi_{\rm ph}^{4}},\qquad
\hat\ell^{2}=P(\theta_{\gamma})\sin^{2}\theta_{\gamma}.
\end{align}
Hence the magnitude of the dimensionless coordinate time orbital frequency $\Omega_{\rm orb}:=|d\phi/d\tau|$ is

\begin{equation}\label{eq:Omega_orb}
\Omega_{\rm orb}
=\frac{\sqrt{\Delta\,\bigl(\xi_{\rm ph}\bigr)}}
       {\xi_{\rm ph}^{2}\, C\sqrt{P(\theta_{\gamma})}\,\sin\theta_{\gamma}},
\end{equation}
The sign of $\Omega_{\rm orb}$ is $\mathrm{sgn}(\hat\ell)$ and fixes the azimuthal orientation. (the reparametrisation $\lambda\leftrightarrow\lambda_{K}$ cancels in
$d\phi/d\tau$). In the Schwarzschild limit $a\to0$,
$\Omega_{\rm orb}\to 1/(3C\sqrt{3})$.

\subsubsection{Lyapunov exponent.}

In the $K$-scaled Mino system the radial equation is
\begin{equation}
\left(\frac{d\xi}{d\lambda_K}\right)^{2}=R(\xi),
\qquad
R(\xi):=\xi^{4}\hat e^{2}-\Delta(\xi),
\end{equation}
cf.\ Equation~\eqref{eq:Rquartic}. For an unstable spherical photon orbit
$\xi=\xi_{\rm ph}$ one has a double root,
$R(\xi_{\rm ph})=0=R'(\xi_{\rm ph})$.
Perturbing $\xi=\xi_{\rm ph}+\delta\xi$ gives
$R(\xi)=\tfrac12 R''(\xi_{\rm ph})(\delta\xi)^2+\cdots$ and hence
\begin{equation}
\delta\xi''=\tfrac12\,R''(\xi_{\rm ph})\,\delta\xi,
\end{equation}
where primes now denote derivatives with respect to $\lambda_K$.
Converting to coordinate time $\tau$ using
\begin{equation}
\left.\frac{d\tau}{d\lambda_K}\right|_{\xi_{\rm ph}}
=\frac{\xi_{\rm ph}^{4}\hat e}{\Delta(\xi_{\rm ph})}
\end{equation}
yields the (dimensionless) Lyapunov exponent
\begin{align}
\Lambda
&=\sqrt{\frac{R''(\xi_{\rm ph})}{\,2\,(d\tau/d\lambda_K)^{2}\,}}\nonumber\\[5pt]
&= \frac{\Delta(\xi_{\rm ph})}{\xi_{\rm ph}^{4}}
\sqrt{\frac{1}{2}\left[12\,\xi_{\rm ph}^{2}
-\frac{\xi_{\rm ph}^{4}}{\Delta(\xi_{\rm ph})}\,\Delta''(\xi_{\rm ph})\right]}\; .
\label{eq:Lyapunov_general}
\end{align}
with $\Delta''(\xi)= -12a^{2}\xi^{2}+12a^{2}\xi+2$.
For $a=0$ one obtains $\Lambda=1/(3\sqrt{3})$, as expected.

\paragraph*{Small-$a$ expansions.}
Using $\xi_{\rm ph}=3\,[1-3a^{2}+\mathcal O(a^{4})]$,
$\cos\theta_{\gamma}=a-3a^{3}+\mathcal O(a^{5})$,
$\sin\theta_{\gamma}=1-\tfrac12 a^{2}+\mathcal O(a^{4})$,
and $P(\theta_{\gamma})=1+2a^{2}+\mathcal O(a^{4})$,
one finds
\begin{subequations}
\begin{align}
\Omega_{\rm orb}
&=\frac{1}{3C\sqrt{3}}\Bigl[\,1-5a^{2}+\mathcal O(a^{4})\Bigr],\\
\Lambda&=\frac{1}{3\sqrt{3}}\Bigl[\,1-\tfrac{3}{2}a^{2}
   +\mathcal O(a^{4})\Bigr].
\end{align}    
\end{subequations}
Thus, acceleration reduces both the pattern speed and the instability rate at quadratic order in $a$; only the azimuthal frequency carries the explicit $1/C$ inherited from $g_{\phi\phi}\propto C^{2}$.

\subsubsection{Eikonal quasinormal estimate and conventions.}

Linear perturbations (scalar, electromagnetic, gravitational) on a stationary
black hole background satisfy wave equations of the schematic form
\begin{equation}
\partial_{\tau}^{2}\Psi-\partial_{r_*}^{2}\Psi+\mathcal V_{\mathfrak m}(x;\omega)\,\Psi=0,    
\end{equation}
where $\tau=t/m$ is our dimensionless coordinate time, $r_*$ is a tortoise type coordinate, and $\mathfrak m$ is the azimuthal number (in spherical cases one would also have $l$, but in the static C-metric axisymmetry singles out $\mathfrak m$). In the eikonal (geometric optics) limit (see e.g.,\cite{PhysRevD.79.064016,2020PhRvD.102d4005D}) $\mathfrak m\gg1$ the wavelength is short compared to curvature scales, and a
WKB ansatz
\begin{equation}
\Psi(\tau,x)\sim A(x)\,\exp\bigl[i\,\mathfrak m\,S(\tau,x)\bigr]
\end{equation}
reduces, at leading order, to the Hamilton-Jacobi equation
$g^{\mu\nu}\partial_{\mu}S\,\partial_{\nu}S=0$, i.e.\ wavefronts propagate along null geodesics. A wave packet initially localised near the unstable photon surface stays trapped for many cycles; its pattern speed is the geodesic angular frequency while the decay rate is controlled by the local instability (Lyapunov exponent) of neighbouring null rays.

The standard matching across the potential peak at $\xi=\xi_{\rm ph}(a)$ (with purely ingoing/outgoing boundary conditions at the two Killing horizons in the static patch) yields the eikonal spectrum (see, e.g., \cite{PhysRevD.79.064016,2020PhRvD.102d4005D})
\begin{equation}\label{eq:QNM-eikonal}
\omega_{\mathfrak m n}
\approx \mathfrak m\,\Omega_{\rm orb}
 - i\Bigl(n+\tfrac12\Bigr)\Lambda , \qquad
\mathfrak m\gg1,\ \ n=0,1,2,\ldots,
\end{equation}
with $\Omega_{\rm orb}$ and $\Lambda$ as in Equations \eqref{eq:Omega_orb}-\eqref{eq:Lyapunov_general}, measured with respect to
$\tau=t/m$; the physical frequency is
$\omega_{\mathfrak m n}^{\rm phys}=\omega_{\mathfrak m n}/m$. Note that $\Lambda$ carries no explicit $C$ factor, whereas
$\Omega_{\rm orb}$ contains an overall $1/C$ coming from $g_{\phi\phi}\propto C^{2}$.

For the static C-metric the unstable spherical photon orbits sit at the single
radius $\xi=\xi_{\rm ph}(a)$ and the unique photon circle at the cone latitude $\theta=\theta_{\gamma}(a)$ (section~\ref{sec:photon-cone}). In the radial WKB problem the effective potential $\mathcal V_{\mathfrak m}$ develops a sharp maximum at $\xi=\xi_{\rm ph}(a)$.\footnote{Formally, one introduces a tortoise coordinate adapted to the spherical orbit, $dr_*/d\xi>0$, e.g.\ the standard choice along $\theta=\theta_\gamma$, so that the potential peak sits at a finite $r_*$. The precise definition of $r_*$ is irrelevant for the leading eikonal result.} A 1D WKB expansion about this maximum, matched to purely ingoing (outgoing) boundary conditions at the black hole (acceleration) horizon in the static patch, yields a quantisation condition whose leading real and imaginary parts are set by classical photon ring data (see, e.g., \cite{PhysRevD.79.064016,2020PhRvD.102d4005D}).

Because the static C-metric breaks equatorial symmetry through
$\theta_{\gamma}(a)$ and $P(\theta_{\gamma})$, the eikonal spectrum depends on
the acceleration $a$ and on the cone latitude via
Equations \eqref{eq:Omega_orb} and \eqref{eq:Lyapunov_general}.

At fixed coordinate harmonic label $\mathfrak m$ (with $\phi\in[0,2\pi)$), since the separated perturbations are $e^{i\mathfrak m\phi}$ with integer $\mathfrak m$, inserting Equation \eqref{eq:Omega_orb} into Equation \eqref{eq:QNM-eikonal}
gives
\begin{equation}
\Re\,\omega_{\mathfrak m n}\ \approx\ 
\frac{\mathfrak m}{C}\,
\frac{\sqrt{\Delta(\xi_{\rm ph})}}
     {\xi_{\rm ph}^{2}\,\sqrt{P(\theta_{\gamma})}\,\sin\theta_{\gamma}},    
\end{equation}
i.e.\ the leading real part scales as $1/C$ when $\mathfrak m$ is held fixed.

It is sometimes convenient to introduce the physical azimuth $\Phi:=C\,\phi$, whose period is $2\pi C$. A separated mode $e^{i\mathfrak m\phi}$ (with integer $\mathfrak m$) can then be written as $e^{i(\mathfrak m/C)\Phi}$. Since
\begin{equation}
\frac{d\Phi}{d\tau}=C\,\Omega_{\rm orb},
\end{equation}
one has the equivalent identity
\begin{equation}
\mathfrak m\,\Omega_{\rm orb}=\frac{\mathfrak m}{C}\,\frac{d\Phi}{d\tau}.
\end{equation}
Thus, $d\Phi/d\tau$ is independent of $C$, while the physical azimuthal wavenumber is rescaled as $\mathfrak m/C$. Therefore,

\begin{itemize}\itemsep2pt
\item The imaginary part is set by $\Lambda$ and is independent of $C$.
\item The real part is $\Re\,\omega\approx \mathfrak m\,\Omega_{\rm orb}
=(\mathfrak m/C)\,(d\Phi/d\tau)$. At fixed integer $\mathfrak m$ it scales as $1/C$; equivalently, at fixed physical wavenumber $\mathfrak m/C$ the leading coefficient is $C$-independent because $d\Phi/d\tau$ contains no $C$.
\end{itemize}

Thus, (i) Equation \eqref{eq:QNM-eikonal} captures the high $\mathfrak m$
ringdown controlled by the unstable photon ring seen by static
observers in the wedge $2<\xi<1/a$. The boundary conditions at the black hole and acceleration horizons fix the quasinormal mode character, but they do not alter the leading eikonal relation above. (ii) The estimate is accurate for $\mathfrak m\gg1$ and low overtones $n=\mathcal O(1)$; subleading corrections (e.g.\ beyond-WKB and spin dependent terms) do not affect the leading identification
$\Re\,\omega=\mathfrak m\,\Omega_{\rm orb}$ and
$\Im\,\omega=-(n+\tfrac12)\Lambda$.

\subsection{Cosmic string tension: what can (and cannot) be constrained}
\label{subsec:string_bound}
The conical parameter $C$ enters only through
$g_{\phi\phi}=\Omega^{-2}P\,C^{2}\,\xi^{2}\sin^{2}\theta$. Locally, the
shadow boundary is obtained from the screen variables  in
Equations \eqref{eq:alpha_general}-\eqref{eq:beta_general}. Combining them gives
\begin{equation}
    \iota^{2}+\vartheta^{2}
=\frac{\Delta_{o}}{\xi_{o}^{4}}\;Q,
\end{equation}
see Equation \eqref{eq:alpha2plusbeta2}. On the photon surface $Q=Q_{\rm ph}$ is fixed by Equation \eqref{eq:Q_spherical}, so the shadow locus satisfies Equation \eqref{eq:shadow_circle}. Crucially, the $C$-dependence cancels between $\iota$ and $\vartheta$, and therefore:

For any static observer, the shadow boundary is a circle with
radius Equation \eqref{eq:Rshadow} that is independent of $C$. Thus, from
the shape and size of the local shadow alone one cannot infer the string tension. In fact, the azimuthal rescaling and the resulting conical defects on the symmetry axes,
$\delta_{\rm N,S}=2\pi\,[\,1-C(1\pm2a)\,]$ (section~\ref{sec:metric}),
modify the global geometry by excising (string, $\delta>0$) or adding (strut, $\delta<0$) a wedge about the axis. This produces global lensing signals (double images separated by the deficit, azimuthal discontinuities across the string, distinct caustic structure) that do not show up in the local shadow boundary. Realistic strategies therefore are:
\begin{itemize}\itemsep1pt
\item infer $a$ from the shadow radius via Equation \eqref{eq:Rshadow} at a known observer position $\xi_{o}$;
\item constrain $C$ from global lensing (multiple images, azimuthal jumps), not from the circle outline of the shadow.
\item Finally, while the shadow boundary itself carries no information about $C$, the eikonal ringdown estimate does: in Section~\ref{sec:ringdown} the Lyapunov exponent $\Lambda$ is $C$-independent, whereas the orbital frequency $\Omega_{\rm orb}$ inherits an overall $1/C$ from $g_{\phi\phi}\propto C^{2}$. Thus, ringdown can in principle provide a complementary on the conicity through the azimuthal sector (subject to the identification of the relevant azimuthal harmonic content).

\end{itemize}
A joint analysis of the bright photon ring (which depends on transport and emissivity) together with global lensing offers a practical path to bounds on $C$.


\section{Classical invariants and diagnostics}
\label{subsec:diagnostics}

Beyond solving individual geodesics, it is useful to collect compact,
coordinate-invariant diagnostics of the static C-metric that (i) provide
quick checks for analytic manipulations and numerical ray tracing, and
(ii) clarify which observables are insensitive to the conical parameter
$C$. In this section we summarise two such diagnostics in the static
patch $2m<r<1/\alpha$ (equivalently $2<\xi<1/a$ with $a=\alpha m$ and
$\Omega>0$): the surface gravities (and associated Killing temperatures)
of the two Killing horizons, and the exact redshift law between any two
static worldlines. Throughout we use the same normalisation of the
stationary Killing vector $\xi_{(t)}=\partial_t$ as in the geodesic
analysis, so the results below are directly comparable across the paper and in the $a\to0$ (Schwarzschild) limit.



\subsection{Surface gravities and thermal (non)equilibrium of horizons}
\label{sec:temps}

As a background diagnostic (and for later comparison with the geodesic time coordinate),
we record the surface gravities of the two Killing horizons in the static patch.
We keep the same Killing normalisation used throughout the paper,
$\xi_{(t)}=\partial_t$, so the resulting $\kappa$ and $T=\kappa/(2\pi)$ are
directly comparable to the geodesic relations and to the Schwarzschild limit.
For thermodynamic discussions of accelerating black holes with conical defects,
see, e.g., \cite{2017JHEP...05..116A,2006CQGra..23.6745G}.
In the static patch the relevant metric components are
\begin{align}
g_{tt}=-\frac{\Delta}{\Omega^{2}\,r^{2}},\qquad
g_{rr}=\frac{r^{2}}{\Omega^{2}\,\Delta},
\end{align}
with
\begin{equation}
\Delta=(r^{2}-2mr)\,(1-\alpha^{2}r^{2}),\qquad
\Omega=1+\alpha r\cos\theta.
\end{equation}
The Killing horizons of $\xi_{(t)}$ occur at the simple zeros of $\Delta(r)$,
namely the black hole horizon $r_{\rm H}=2m$ and the acceleration horizon
$r_{\rm A}=1/\alpha$.

Surface gravity for the canonical Killing field:
For a static metric of the above form, the surface gravity of $\xi_{(t)}$ at a
Killing horizon $r=r_{\rm h}$ can be written as
\begin{equation}
\kappa
=\frac12\,\partial_{r}(-g_{tt})\,
\sqrt{\frac{g^{rr}}{-g_{tt}}}\Big|_{r=r_{\rm h}}
=\frac{1}{2\,r_{\rm h}^{2}}\,\Delta'(r_{\rm h}) .
\label{eq:kappa_general}
\end{equation}
Two points are worth stressing.

\begin{itemize}\itemsep2pt
\item The $\theta$-dependence from $\Omega(r,\theta)$ cancels in Equation \eqref{eq:kappa_general},
so $\kappa$ is constant over each horizon (as required for a Killing horizon).
\item The conicity parameter $C$ enters only through $g_{\phi\phi}$, hence it does not
affect $g_{tt}$ or $g_{rr}$; therefore $\kappa$ (and $T$) are independent of $C$.
\end{itemize}

Differentiating $\Delta$ gives
\begin{equation}
\Delta'(r)=(2r-2m)(1-\alpha^{2}r^{2})-2\alpha^{2}r(r^{2}-2mr).
\end{equation}
Evaluating at the horizons yields
\begin{subequations}
\begin{align}
\Delta'(r_{\rm H})&=2m\bigl(1-4\alpha^{2}m^{2}\bigr),\\
\Delta'(r_{\rm A})&=-2\,(r_{\rm A}-2m).
\end{align}
\end{subequations}
Hence
\begin{equation}
\kappa_{\rm H}=\frac{1-4a^{2}}{4m},\qquad
\kappa_{\rm A}
=-\,\frac{r_{\rm A}-2m}{r_{\rm A}^{2}}=-\,\frac{a-2a^{2}}{m},
\qquad (a:=\alpha m).
\label{eq:kappas}
\end{equation}
We will typically use the positive magnitude for the acceleration horizon,
$|\kappa_{\rm A}|=(a-2a^{2})/m$. So, the associated Killing temperatures are
\begin{equation}
T_{\rm H}=\frac{\kappa_{\rm H}}{2\pi},\qquad
T_{\rm A}=\frac{|\kappa_{\rm A}|}{2\pi},
\end{equation}
with the understanding that in non-asymptotically flat spacetimes the
overall scale depends on the chosen normalisation of $\xi_{(t)}$
(see, e.g., the thermodynamic discussions in
\cite{2017JHEP...05..116A}).
No global thermal equilibrium:
For $0<a<\tfrac12$ one finds
\begin{equation}
\kappa_{\rm H}-|\kappa_{\rm A}|
=\frac{(1-2a)^{2}}{4m}\;>\;0,
\end{equation}
with equality only at the degenerate limit $a\to\tfrac12$, where $r_{\rm H}=r_{\rm A}$
and both surface gravities vanish. Thus, with respect to the same Killing time $t$,
the two horizons generically have different temperatures, and the static patch cannot be
in global thermal equilibrium (Tolman law). Any constant rescaling of $\xi_{(t)}$
rescales $\kappa_{\rm H}$ and $\kappa_{\rm A}$ by the same factor and cannot remove
their mismatch.

Near-horizon Rindler form: Fix a latitude $\theta$ with $\Omega_{\rm h}:=\Omega(r_{\rm h},\theta)>0$ (away from the corner where $\Omega_{\rm h}=0$ on the acceleration horizon). By using $\Delta(r)=\Delta'(r_{\rm h})(r-r_{\rm h})+\mathcal{O}\!\bigl((r-r_{\rm h})^{2}\bigr)$, the $(t,r)$-sector becomes
\begin{equation}
ds^{2}_{(t,r)}
\simeq -\frac{\Delta'(r_{\rm h})(r-r_{\rm h})}{\Omega_{\rm h}^{2}r_{\rm h}^{2}}\,dt^{2}
+\frac{r_{\rm h}^{2}}{\Omega_{\rm h}^{2}\Delta'(r_{\rm h})(r-r_{\rm h})}\,dr^{2}.
\end{equation}
Introducing the proper distance $\rho$ normal to the horizon gives the standard Rindler form
\begin{equation}
ds^{2}\simeq -\kappa^{2}\rho^{2}dt^{2}+d\rho^{2}
+\frac{r_{\rm h}^{2}}{\Omega_{\rm h}^{2}P(\theta)}\,d\theta^{2}
+\frac{P(\theta)\,C^{2}r_{\rm h}^{2}\sin^{2}\theta}{\Omega_{\rm h}^{2}}\,d\phi^{2},
\end{equation}
with $\kappa=|\Delta'(r_{\rm h})|/(2r_{\rm h}^{2})$, consistent with Equation \eqref{eq:kappa_general}.
\footnote{At the corner where $\Omega_{\rm h}=0$ (south pole of the acceleration horizon) this Gaussian normal construction must be modified.}

\subsection{Redshift map for static worldlines}
\label{sec:redshift}

Let a photon with wavevector $k^{\mu}$ be emitted by a static source at
$(\xi_{\rm e},\theta_{\rm e})$ and received by a static observer at
$(\xi_{\rm o},\theta_{\rm o})$, both within the static patch.
A static worldline is an integral curve of $\partial_t$ with unit 4-velocity
\begin{equation}
u^\mu = N^{-1}\,\delta^\mu{}_t,
\qquad
N(\xi,\theta):=\sqrt{-g_{tt}}.
\end{equation}
The measured frequency is $\omega=-k\cdot u$ (see e.g., \cite{Wald1984,Poisson2004}). Since the spacetime is stationary, the Killing energy $E:=-k_t$ is conserved along the photon trajectory, hence for any static observer
\begin{equation}
\omega(\xi,\theta)=\frac{E}{N(\xi,\theta)}.
\end{equation}
Therefore the redshift factor between two static worldlines depends only on the endpoints and is given exactly by
\begin{equation}
\frac{\omega_{\rm o}}{\omega_{\rm e}}
=\frac{N(\xi_{\rm e},\theta_{\rm e})}{N(\xi_{\rm o},\theta_{\rm o})}
=\sqrt{\frac{-g_{tt}(\xi_{\rm e},\theta_{\rm e})}{-g_{tt}(\xi_{\rm o},\theta_{\rm o})}}.
\label{eq:redshift_static_general}
\end{equation}
For the static C-metric,
\begin{equation}
-g_{tt}(\xi,\theta)=\frac{\Delta(\xi)}{\Omega(\xi,\theta)^{2}\,\xi^{2}}
=\frac{\bigl(1-\tfrac{2}{\xi}\bigr)\,\bigl(1-a^{2}\xi^{2}\bigr)}{\Omega(\xi,\theta)^{2}},
\end{equation}
so one may write the redshift factor in the explicit dimensionless form
\begin{equation}
\frac{\omega_{\rm o}}{\omega_{\rm e}}
=\sqrt{\frac{(1-\tfrac{2}{\xi_{\rm e}})\,(1-a^{2}\xi_{\rm e}^{2})}
            {(1-\tfrac{2}{\xi_{\rm o}})\,(1-a^{2}\xi_{\rm o}^{2})}}\;
\frac{\Omega(\xi_{\rm o},\theta_{\rm o})}{\Omega(\xi_{\rm e},\theta_{\rm e})}.
\label{eq:redshift_static}
\end{equation}
This makes three structural points transparent:
(i) the redshift is endpoint-only (no dependence on the intervening null path);
(ii) it is independent of the conicity parameter $C$, since $g_{tt}$ contains no
$C$-factor; and
(iii) it diverges as either endpoint approaches a Killing horizon where
$N\to0$ (no static worldline exists exactly on the horizon).

Schwarzschild limit:
As $a\to0$, one has $\Omega\to1$ and $(1-a^{2}\xi^{2})\to1$, so
Equation~\eqref{eq:redshift_static} reduces to the standard Schwarzschild
static redshift
\begin{equation}
\frac{\omega_{\rm o}}{\omega_{\rm e}}
\to \sqrt{\frac{1-\tfrac{2}{\xi_{\rm e}}}{1-\tfrac{2}{\xi_{\rm o}}}}.
\end{equation}

Small-$a$ anisotropy:
At fixed $(\xi_{\rm e},\theta_{\rm e})$, $(\xi_{\rm o},\theta_{\rm o})$, the only
\emph{linear} dependence on $a$ comes from $\Omega(\xi,\theta)=1+a\xi\cos\theta$,
since $\Delta(\xi)$ depends on $a$ only through $a^{2}$. Expanding
Equation \eqref{eq:redshift_static} gives
\begin{equation}
\frac{\omega_{\rm o}}{\omega_{\rm e}}
= \sqrt{\frac{1-\tfrac{2}{\xi_{\rm e}}}{1-\tfrac{2}{\xi_{\rm o}}}}\,
\Bigl[\,1+a\bigl(\xi_{\rm o}\cos\theta_{\rm o}-\xi_{\rm e}\cos\theta_{\rm e}\bigr)
+\mathcal O(a^{2})\Bigr],
\end{equation}
i.e.\ the usual Schwarzschild redshift multiplied by an anisotropic
north--south factor. In particular, at equal radii
$\xi_{\rm e}=\xi_{\rm o}$ one finds
$\omega_{\rm o}/\omega_{\rm e}=\Omega(\xi,\theta_{\rm o})/\Omega(\xi,\theta_{\rm e})$,
so even co-accelerated static observers at the same $\xi$ but different
latitudes see a nontrivial accelerative redshift.

Use as a diagnostic: Equation~\eqref{eq:redshift_static} is a simple invariant check for analytic and numerical work: for ray tracing with static emitters/observers the measured frequency must obey $\omega=E/N$, and the ratio between any two static worldlines must match Equation \eqref{eq:redshift_static} independently of the photon path. This also cleanly separates effects due to acceleration (through $\Omega$ and the
factor $1-a^{2}\xi^{2}$) from effects due to the conical deficit (absent here but
present in global lensing through the azimuthal geometry).



\section{Summary and Conclusions}
\label{sec:conclusions}

We analysed null geodesics and photon observables in the subextremal static C-metric, working in a fully dimensionless formulation. Introducing a Mino-type parameter and a convenient $K$-scaling for photons, we obtained a uniform first-order system whose reduced dynamics is controlled by two dimensionless ratios, allowing a parameter-consistent treatment across all sectors.

Because null motion is conformally invariant, the Hamilton-Jacobi equation separates. We reduced both the radial and polar sectors to Biermann-Weierstrass form and expressed the general solutions in terms of $\wp$, $\zeta$, and $\sigma$. This yields a clean classification of photon trajectories via the radial quartic and polar potential, makes the breakdown of equatorial symmetry explicit through the existence of a fixed photon cone at nonzero acceleration, and identifies a single spherical photon surface shared by all latitudes. For generic trajectories, the azimuthal and time variables are obtained as closed Weierstrass quadratures built from the independent radial and polar phases, while cone orbits collapse to a single phase and admit particularly simple behaviour.

On the observational side, we proved that the black hole shadow boundary is an exact circle for any observer in the static patch. For static observers, its angular radius depends on the acceleration and the observer location but is independent of observer inclination and of the conicity parameter $C$, so the local shadow alone cannot constrain the string/strut tension. Moreover, the circular shadow radius leads to a simple algebraic inference scheme that determines the dimensionless acceleration from a single shadow measurement once the observer’s mass-distance relation is fixed. We further clarified the observer dependence of the shadow scale: at a fixed event, changing the observer’s state of motion acts by local aberration, which preserves circularity but changes the observed angular size, so the shadow radius is not invariant under boosts.

We also provided compact expressions for the photon orbital frequency and Lyapunov exponent and used the geodesic-QNM correspondence to obtain eikonal quasinormal estimates, showing that the damping rate is insensitive to the conical deficit at this order, while the oscillation frequency is affected by $C$ only through a simple rescaling of the azimuthal mode number.

As background diagnostics, we computed the surface gravities (and hence temperatures) of the black hole and acceleration horizons and showed that they are generically unequal, precluding global thermal equilibrium with respect to the stationary Killing time. We also derived an exact redshift law between static worldlines throughout the static patch. Both diagnostics are independent of $C$ and provide practical consistency checks for analytic calculations and numerical ray tracing.

Overall, acceleration modifies photon dynamics in a transparent way by shifting the photon surface and replacing equatorial symmetry by a distinguished photon cone, while the shadow remains exactly circular with an acceleration-dependent angular scale set jointly by the photon surface and the observer location. The conical deficit influences global lensing and image multiplicities, but not the local shadow boundary. Taken together, these results provide a unified analytic framework for photon dynamics and observables in the static C-metric and suggest concrete inference strategies: extract acceleration from the shadow radius (for a known observer configuration) and constrain the string tension from global lensing signatures rather than from the local shadow.

Several extensions are natural: (i) include electric charge and track the deformation of the photon cone, photon surface, and ringdown data; (ii) incorporate slow rotation to study the interplay of frame dragging and translatory acceleration in the shadow and eikonal spectrum; (iii) couple the analytic geodesic solutions to radiative transfer and/or plasma refraction to predict intensity maps beyond the geometric boundary; (iv) develop a global lensing analysis to translate large scale image distortions into quantitative bounds on the conical deficit; (v) also monte carlo study of accretion onto this black holes ( see e.g., \cite{2023PhRvD.108l4057M,2024PhRvD.110h4014C}) and (v) benchmark the eikonal estimates against full perturbation theory in the static C-metric wedge to quantify subleading corrections and boundary condition effects at the acceleration horizon.


\acknowledgments
The authors are grateful to Vladim\'{\i}r Karas for his worthwhile support in the initial stage of this work, and Avery E. Broderick for invaluable comments and discussions. S.F. acknowledges the University of Waterloo, Natural Sciences and Engineering Research Council of Canada, of the Government of Canada through the Department of Innovation, Science and Economic Development and the Province of Ontario through the Ministry of Colleges and Universities at Perimeter Institute, and ZARM institute in Germany. E.H. acknowledges support by the Deutsche Forschungsgemeinschaft (DFG, German Research Foundation) under Germany’s Excellence Strategy-EXC-2123 QuantumFrontiers-390837967. 

%

\bibliography{cmetricsecond}
\bibliographystyle{unsrt}

\appendix
\section{Weierstrass toolkit for null geodesics}
\label{app:null:weierstrass}

For completeness, we collect the explicit variable changes and $\sigma,\zeta$ integrals underlying section \ref{sec:null-dimless}, using the dimensionless notation Equation \eqref{eq:dimless_defs}
$a=\alpha m$, $\varepsilon=mE$, $\ell=L/m$,
$\ell_C=\ell/C$, and $\xi=r/m$.

\subsection{Weierstrass reduction}
The first order null equations in Mino gauge are (see Equation \eqref{eq:firstorder_dimless} and Equation \eqref{eq:polar_nu_lamK}),
\begin{subequations}
\begin{align}
&\left(\frac{d\xi}{d\lambda_{K}}\right)^{2}
= R(\xi) = \xi^{4}\,\hat e^{2}-\Delta(\xi),
\\[3pt]
&\left(\frac{d\nu}{d\lambda_{K}}\right)^{2}
= \Theta_\nu(\nu) = (1-\nu^2)\,P(\nu)-\hat\ell^{2},\\
&\quad \frac{d\phi}{d\lambda_{K}}
=\frac{\hat\ell}{CP(\nu)\,(1-\nu^2)},
\qquad
\frac{d\tau}{d\lambda_{K}}
=\frac{\xi^{4}\,\hat e}{\Delta(\xi)}. \label{app:eq:phit}
\end{align}
\end{subequations}
where $\nu=\cos\theta$ and $P(\nu)=1+2a\nu$. For the radial and polar sector, the general strategy is to reduce the problem to the Weierstrass cubic $(dx/d\lambda_K)^2 = P_W(x) = 4x^3-g_2x-g_3$ with invariants $g_2$, $g_3$. If the roots of the polynomials on the right hand side are known, this can be achieved by a direct substitution. If the roots are not known or their calculation should be avoided, a generalised approach by Biermann can be used.

\mycomment{
\begin{align}
&\Bigl(\tfrac{d\xi}{d\lambda}\Bigr)^{2}
 = R(\xi)
  = \varepsilon^{2}\xi^{4}-\Delta(\xi)\,K,\\
  & \Delta(\xi)=\xi^{2}\Bigl(1-\tfrac{2}{\xi}\Bigr)(1-a^{2}\xi^{2}),\\
&\Bigl(\tfrac{d\theta}{d\lambda}\Bigr)^{2}
 = \Theta(\theta)
  = P(\theta) K-\frac{\ell_{C}^{2}}{\sin^{2}\theta},\\
 &P(\theta)=1+2a\cos\theta .
\end{align}
}

\paragraph{Radial and polar sector.}
\mycomment{
The classical approach to solve the equation is as follows:
Set $u=1/\xi$ and then $u=y+\tfrac{1}{6}$, followed by
$Y=\tfrac{K}{2}y$. One obtains the standard Weierstrass form
\begin{align}
&\Bigl(\tfrac{dY}{d\lambda}\Bigr)^{2}
 = 4Y^{3}-g_{2}^{(r)}\,Y-g_{3}^{(r)},\\
& g_{2}^{(r)}=K^{2}\Bigl(a^{2}+\tfrac{1}{12}\Bigr),\\
& g_{3}^{(r)}=\frac{K^{3}}{216}-\frac{K^{3}a^{2}}{6}-\frac{K^{2}\varepsilon^{2}}{4}.
\end{align}
Hence $Y(\lambda)=\wp(\lambda-\lambda_{r0};g_{2}^{(r)},g_{3}^{(r)})$ and
\begin{equation}
\xi(\lambda)=\frac{1}{\tfrac{2}{K}\,Y(\lambda)+\tfrac{1}{6}}
=\frac{1}{\tfrac{2}{K}\,\wp(\lambda-\lambda_{r0})
        +\tfrac{1}{6}} . \label{app:eq:xi}
\end{equation}
}
The Biermann-Weierstrass approach uses Taylor's theorem,
\begin{equation}\label{app:taylor}
    R(\xi) = \sum_{n=0}^4 \frac{1}{n!} \frac{d^nR}{d\xi^n}(\xi_0) (\xi-\xi_0)^n\,.
\end{equation}
If $\xi_0$ is a zero of $R$, then it is straightforward to find that a substitution $\xi=\xi_0+\frac{R_0'}{4y-R_0''/6}$ leads to the Weierstrass cubic with invariants
\begin{subequations}
\begin{align}
    g_2 & = \frac{1}{48} (R_0''-2R_0'R_0'''), \\
    g_3 & = \frac{1}{12^3} \left( 48 R_0'R_0'' R_0''' - R_0''^3 - 29 \cdot 12^2 R_0'^2 R_0''''\right)\,.
\end{align}
\end{subequations}
Here the index $0$ indicates evaluation at $\xi_0$. These invariants do actually not dependent on the choice of $\xi_0$ and can be expressed directly in terms of the coefficients of $R$ as
\begin{align} \label{app:g2g3}
    g_2 & = b_0b_4 - \frac{1}{4} b_1b_3 + \frac{1}{12} b_2^2\,\\
    g_3 & = \frac{1}{6}b_0 b_2 b_4 -\frac{1}{16} b_0 b_3^2 - \frac{1}{216} b_2^3- \frac{1}{16} b_1^2 b_4 + \frac{1}{48} b_1 b_2 b_3,\nonumber
\end{align}
where $R(\xi) = \sum_{n=0}^4 b_n \xi^n$. 

\mycomment{
Note that the solution Equation \eqref{app:eq:xi} is of course equivalent to the solution given in the main body of the paper. This is due to the fact that for orbital motion that does not reach $r=0$ the argument of the Weierstrass function in Equation \eqref{app:eq:xi} is complex. It is given by $\lambda - \lambda_{r0} = x + \omega_2$, where $x$ is some real number and $\omega_2$ is the complex half-period of $\wp$. Using the addition theorems for $\wp$ the equivalence to Equation \eqref{eq:xi_r0_solution} can be explicitly demonstrated.
}

The above approach can be generalized to any arbitrary initial value $\xi_0$. For details of the derivation, we refer to \cite{Cieslik2022} (Appendix A) and references therein. The idea is to introduce an auxiliary function $w(\xi;\xi_0)$ such that $(dw/d\xi)^2 = P_W(w)/R(\xi)$ and, therefore, $w = \wp(\lambda_K)$ with the invariants as in Equation \eqref{app:g2g3}. The function $w$ can be calculated explicitly (see eq.~(A.21) in \cite{Cieslik2022}) and be given in various forms including
\begin{align}
    w = \frac{F_1(\xi,\xi_0) + \sqrt{R(\xi) R_0}}{2(\xi-\xi_0)^2}
\end{align}
with $F_1(\xi,\xi_0) = R_0+R_0'(\xi-\xi_0)/2+R_0''(\xi-\xi_0)^2/12$. Note that $w$ is related to the global Abelian 2-differential of the second kind with the unique pole of order 2 along $\xi=\xi_0$ . It solves a quadratic equation 
\begin{align}
    (\xi-\xi_0)^2w^2 - F_1(\xi,\xi_0)w + F_2(\xi,\xi_0) = 0 
\end{align}
with $4(\xi-\xi_0)^2 F_2(\xi,\xi_0)=F_1(\xi,\xi_0)^2 - R(\xi)R_0$. Using again the Taylor series \eqref{app:taylor}, the last equation can be rewritten as a quadratic equation in $(\xi-\xi_0)$ with solution
\begin{align}
    \xi-\xi_0 = \frac{R_0'(w-\frac{1}{24}R_0'')+\frac{1}{12}R_0R_0'''-2\epsilon_r\sqrt{R_0 P_W(w)}}{4(w-\frac{1}{24}R_0'')^2 - \frac{1}{24}R_0R_0''''}
\end{align}
which directly gives the final result Equation \eqref{eq:solr}.

The procedure for the polar sector is completely analogous.

\mycomment{
\paragraph{Polar sector.}
With $\nu=\cos\theta$, define $Y_{\theta}$ by
$\nu=-(2Y_{\theta}+\tfrac{1}{6})/a$ and $\gamma=\sqrt{K}\lambda$.
Then
\begin{align}
&\Bigl(\tfrac{dY_{\theta}}{d\gamma}\Bigr)^{2}
 = 4Y_{\theta}^{3}-g_{2}^{(\theta)}\,Y_{\theta}-g_{3}^{(\theta)},
\\
&g_{2}^{(\theta)}=\frac{1}{12}+a^{2},\\
&g_{3}^{(\theta)}=\frac{1}{216}
 +a^{2}\,\Bigl(\frac{\ell_{C}^{2}}{4K}-\frac{1}{6}\Bigr),
\end{align}
hence $Y_{\theta}(\gamma)=\wp(\gamma-\gamma_{0};g_{2}^{(\theta)},g_{3}^{(\theta)})$ and
\begin{equation}
\theta(\lambda)=\arccos\,\left[-\frac{1}{a}\Bigl(2Y_{\theta}(\sqrt{K}\lambda)+\tfrac{1}{6}\Bigr)\right].
\end{equation}
}
\subsection{Azimuth and time integrals}
For the azimuth and coordinate time, eq.~\eqref{app:eq:phit}, partial fraction decompositions produce
\begin{align}
    \frac{d\phi}{d\lambda_K} & = \frac{\hat \ell}{C} \sum_{i=1}^3 \frac{\beta_i}{\nu-c_i} \,,\\
    \frac{d\tau}{d\lambda_K} & = \frac{\hat{e}}{a^2} \left[ -1 + \sum_{i=1}^3 \frac{\gamma_i}{\xi-h_i} \right], \label{app:eq:t}
\end{align}
with the poles
\begin{align}
    c_1 & = 1, \quad c_2 = -1, \quad c_3 = -\frac{1}{2a},\\
    \gamma_1 & = \frac{1}{2a(2a-1)}, \quad \gamma_2 = \frac{1}{2a(2a+1)}, \quad \gamma_3 = \frac{8a^2}{1-4a^2},
\end{align}
and the coefficients
\begin{align}
   \beta_1 & = \frac{-1}{2+4a}, \quad \beta_2 = \frac{1}{2-4a}, \quad \beta_3 = \frac{2a}{(4a^2-1)} \,,\\
   h_1 & = \frac{1}{a}, \quad h_2 = - \frac{1}{a}, \quad h_3 = 2\,.
\end{align}
The integration of azimuth and coordinate time, therefore, reduces to the integration of primitives of the form 
\begin{align}
    p(\lambda_K) := \frac{1}{x(\lambda_K)-c}
\end{align}
with $x=\nu$ or $x=\xi$ and some constant $c$. As $x$ is an elliptiuc function, $p$ is elliptic as well and can be written in terms of Weierstrass functions. We find
\begin{align}
    \frac{dp}{d\lambda_K} & = -p^2 \epsilon_x \sqrt{X(x(p))}\\
    & = - \epsilon_x \left[ \sum_{n=0}^4 \frac{1}{n!} \frac{d^nX}{dx^n}(c) \, p^{(4-n)} \right]^{\frac12}\\
    & =: -\epsilon_x \sqrt{P(p)}
\end{align}
where $X=R$ or $X=\Theta_\nu$ is the right hand side of the radial or polar sector, respectively, and $\epsilon_x = \pm 1$ encodes the initial direction of $x$. 

The invariants for $p$ are identical to the radial and polar equations, respectively, as can be easily checked and $p(\lambda_K)$ becomes
\begin{align}
    p-p_0 = \frac{P_0'(\wp-\frac{1}{24}P_0'')+\frac{1}{12}P_0P_0'''+2\epsilon_x\sqrt{P_0}\wp'}{4(\wp-\frac{1}{24}P_0'')^2 - \frac{1}{24}P_0P_0''''}\,. \label{app:eq:solp}
\end{align}
Here the index $0$ indicates evaluation at the initial value $p_0 = \frac{1}{x_0-c}$, and for notational easy we dropped the argument $\lambda_K$ and the invariants $g_2$, $g_3$ of the Weierstrass functions.

In the main body of the paper we used the simplified version of eq.~\eqref{app:eq:solp} that arises if we assume that $\nu_0$ and $\xi_0$ are turning points, resulting in $P_0=0$. Then the integral of $p$ is
\begin{align}
    \int p(\lambda_K) d\lambda_K & = p_0 \lambda_K + \frac{P_0'}{4} \int \frac{d\lambda_K}{\wp(\lambda_K) - P_0''/24}\\
       & = p_0\lambda_K + \frac{1}{4} P_0' \, \mathcal{F}^{(x)}_{P_0''/24}(\lambda_K)\,,
\end{align}
where the antiderivative $\mathcal{F}$ is given in eq.~\eqref{eq:Weier_primitive}. Substituting back $p_0$ in terms of $r_0$ or $\nu_0$ and expressing the derivatives of $P$ at $p_0$ in terms of $R$ or $\Theta_\nu$, respectively, gives the solutions \eqref{eq:phi_solution} and \eqref{eq:tau_solution} in the main body of the paper.

It is also possible to integrate eq.~\eqref{app:eq:solp} for arbitrary $r_0$ and $\nu_0$. To achieve this, we rewrite eq.~\eqref{app:eq:solp} as
\begin{align}
    p-p_0 = \frac{P_0'\wp+\frac{1}{24}(2P_0P_0''-P_0'P_0'')+2\epsilon_x\sqrt{P_0}\wp'}{4(\wp-P_+)(\wp-P_-)}\,,
\end{align}
where $P_\pm = \frac{1}{24} P_0'' \pm \sqrt{\frac{1}{96}P_0P_0''''}$. The integral of $p$ then is
\begin{align}
    \int p(\lambda_K) d\lambda_K & =  p_0 \lambda_K + \frac{2P_0 P_0''-P_0'P_0''}{96} \mathcal{G}^{(x)}_{P_\pm}(\lambda_K)\\ \nonumber
    & \quad + \frac{P_0'}{4} \mathcal{H}^{(x)}_{P_\pm}(\lambda_K) + \frac{\epsilon_x \sqrt{P_0}}{2} \mathcal{I}^{(x)}_{P_\pm}(\lambda_K),
\end{align}
with the integrals 
\begin{align}
    \mathcal{G}^{(x)}_{P_\pm}(z) & 
    = \frac{\mathcal{F}^{(x)}_{P_+}(z) - \mathcal{F}^{(x)}_{P_-}(z)}{P_+-P_-}\,,\\
    \mathcal{H}^{(x)}_{P_\pm}(z) & 
    = \frac{P_+\mathcal{F}^{(x)}_{P_+}(z) - P_-\mathcal{F}^{(x)}_{P_-}(z)}{P_+-P_-}\,,\\
    \mathcal{I}^{(x)}_{P_\pm}(z) & 
    = \frac{\ln(\wp(z)-P_+) - \ln(\wp(z)-P_-)}{P_+-P_-}\,.
\end{align}

\mycomment{
\begin{equation}
c_1=\frac{a}{2}-\frac{1}{12},\qquad
c_2=-\frac{a}{2}-\frac{1}{12},\qquad
c_3=\frac{1}{6},
\end{equation}
and, for each $c$, a pair of preimages $z_{1},z_{2}$ with
$\wp(z_{s})=c$. Define
\begin{equation}
f(z;z_{1},z_{2})
=\sum_{s=1}^{2}\frac{
 \log\sigma(z-z_{s})-\log\sigma(z_{0}-z_{s})
 -\zeta(z_{s})(z-z_{0})}{\wp'(z_{s})},
\end{equation}
where $\sigma,\zeta$ are the standard Weierstrass functions and $z_0$ is the initial argument (from the radial or polar sector as indicated below). The coefficients
\begin{align}
&\beta_{i}=\Bigl\{\tfrac{1}{4(2a-1)},\;\tfrac{1}{4(2a+1)},\;-\tfrac{a}{4a^{2}-1}\Bigr\},\\
&\gamma_{i}=\Bigl\{-\tfrac{1}{4a(2a-1)},\;-\tfrac{1}{4a(2a+1)},\;\tfrac{1}{4a^{2}-1}\Bigr\}
\end{align}
yield the compact expressions
\begin{align}
\phi(\lambda)&=\phi_{0}
 +\frac{a\,\ell}{C^{2}}
  \sum_{i=1}^{3}\beta_{i}\,
  f\bigl(\sqrt{K}\lambda-\gamma_{0};z_{i1},z_{i2}\bigr),
\\
\tau(\lambda)&=\tau_{0}
 +\varepsilon
  \sum_{i=1}^{3}\gamma_{i}\,
  f\bigl(\lambda-\lambda_{r0};z_{i1},z_{i2}\bigr).
\end{align}

For cone orbits ($\theta=\theta_{\gamma}$ fixed) the arguments coincide and both $\phi$ and $\tau$ are functions of a single Weierstrass phase; for generic C–metric motion the radial and polar arguments differ (hyperelliptic structure), but the $\sigma,\zeta$ representation remains valid.
}


\end{document}